\documentclass[twocolumn,prd,superscriptaddress,preprintnumbers,nofootinbib]{revtex4-1}[11pt]
\usepackage{amsmath,amssymb,graphicx}
\graphicspath{{figs/}}
\usepackage{epsf,verbatim}
\usepackage{hyperref}
\usepackage{comment}
\usepackage{color}
\usepackage{slashed}
\usepackage{subfigure}
\usepackage{comment}
\usepackage{mdwlist, paralist}
\usepackage{rotating}
\usepackage{bm}
\usepackage{multirow}
\usepackage{lipsum}
\usepackage{fontawesome5}
\usepackage{orcidlink}

\newcommand{\beq}{\begin{equation}}
\newcommand{\eeq}{\end{equation}}
\newcommand{\bal}{\begin{align}}
\newcommand{\eal}{\end{align}}
\newcommand{\bit}{\begin{itemize}}
\newcommand{\eit}{\end{itemize}}
\newcommand{\ben}{\begin{enumerate}}
\newcommand{\een}{\end{enumerate}}

\renewcommand{\eqref}[1]{Eq.~(\ref{#1})}

\newcommand{\f}{\frac}

\begin{document}     

\title{Black Hole Cold Brew: Fermi Degeneracy Pressure}

\author{Wei-Xiang Feng\,\orcidlink{0000-0002-9048-2992}}
\email{wxfeng@mail.tsinghua.edu.cn}
\affiliation{Department of Physics, Tsinghua University, Beijing 100084, China}

\author{Hai-Bo Yu\,\orcidlink{0000-0002-8421-8597}}
\email{haiboyu@ucr.edu}
\affiliation{Department of Physics and Astronomy, University of California, Riverside, CA 92521, USA}

\author{Yi-Ming Zhong\,\orcidlink{0000-0001-9922-6162}}
\email{yiming.zhong@cityu.edu.hk}
\affiliation{Department of Physics, City University of Hong Kong, Kowloon, Hong Kong SAR, China}

%\date{\today}

\begin{abstract}
We investigate the dynamical instability of a self-gravitating thermal system in the quantum regime, where Fermi degeneracy pressure becomes significant. Using a truncated Fermi–Dirac distribution and solving the Tolman–Oppenheimer–Volkoff equation, we identify marginally stable configurations following Chandrasekhar’s criterion. While Fermi pressure stabilizes a system against gravitational collapse in Newtonian gravity, in general relativity it can instead drive the instability, enabling collapse even at low  temperatures. In the low-temperature limit, the critical mass is independent of the boundary temperature. We discuss implications for the formation of massive black holes in the early Universe through the gravothermal collapse of dark matter.

\end{abstract}

\maketitle

\section{Introduction} 

Black holes are an inevitable consequence of gravitational collapse as predicted by Einstein’s general relativity, and their existence have been confirmed in various astrophysical observations, including those associated with quasars~\cite{Mortlock:2011wz,Wu:2015vd,Banados:2017unc,Wang:2018tr,Onoue:2019ty,Matsuoka:2019vy,Yang:2020ut,Wang:2021ve,Larson:2023aa}, stellar dynamics~\cite{Schodel:2002py,Ghez:2008ms,GRAVITY:2018ofz}, gravitational-wave detections, and the James Webb Space Telescope (JWST)~\cite{Matthee:2023utn,Harikane:2023aa,Kocevski:2023aa,Pacucci:2023oci,Maiolino:2023bpi}. In theory, stellar-mass black holes, such as those detected by LIGO via merger events, typically form from the death of massive stars~\cite{LIGOScientific:2016aoc,Olsen:2022pin,KAGRA:2021duu,LIGOScientific:2025rid}. The origin of supermassive black holes with masses larger than $10^6{\rm\,M}_\odot$, particularly those in the early Universe, is more puzzling (see~\cite{Inayoshi:2019fun,Woods:2018lty,Haemmerle:2020iqg,Volonteri:2021sfo} for reviews and references therein). A popular mechanism is the direct collapse of pristine gas in proto-galaxies into a massive seed black hole, without first undergoing star formation~\cite{Rees:1984si,Bromm:2002hb,Begelman:2006db}, and the seed further grows by accretion. Primordial black holes~\cite{Carr:2025kdk,Bird:2016dcv,Sasaki:2016jop,Jedamzik:2020omx,Carr:2020gox,Musco:2023dak,Andres-Carcasona:2024wqk,Feng:2024obn} could also play a role in seeding supermassive black holes in the early Universe~\cite{Davoudiasl:2021ijv,Flores:2023zpf,Lu:2024xnb,Qin:2025ymc,Zhang:2025tgm}. Another intriguing mechanism is that if dark matter particles carry self-interactions~\cite{Tulin:2017ara}, the central regions of dark matter halos can undergo gravothermal collapse and form a seed black hole, such interactions could be elastic~\cite{Balberg:2001qg,Pollack:2014rja,Feng:2020kxv,Meshveliani:2022rih,Jiang:2025jtr,Roberts:2025poo,Feng:2025rzf} or inelastic~\cite{Choquette:2018lvq,Xiao:2021ftk,Fernandez:2022zmc,Shen:2025evo}.   

It is important to investigate the dynamical conditions under which a seed black hole can form in light of these seeding mechanisms. The dynamical instability of self-gravitating systems within the framework of general relativity was first studied in the context of supermassive stars~\cite{Chandrasekhar:1964zza,Chandrasekhar:1964zzb,Fowler:1964zza,Bardeen:1965,Zeldovich:1966,Ipser:1980}. In general, a star tends to collapse into a black hole once its gravitational energy becomes comparable to its rest-mass energy, i.e., $GM^2/R \sim Mc^2$, where $G$ is Newton’s constant, $c$ is the speed of light, and $R$ is the characteristic radius of the star. In contrast to Newtonian gravity, where pressure acts as a ``force'' opposing collapse, in general relativity pressure is part of the energy-momentum tensor and therefore sources gravity, which in turn destabilizes the star. Additional factors, such as the equation of state, the cosmological constant and spatial dimensionality~\cite{Boehmer:2005kk,Bordbar:2015wva,Posada:2020svn,Feng:2021acd,Feng:2022fuk}, also play a role in determining stability.

In Ref.~\cite{Feng:2021rst}, we modeled a gravothermal sphere composed of ideal gas particles using a truncated Maxwell–Boltzmann distribution and analyzed its dynamical instability within general relativity. After solving the Tolman--Oppenheimer--Volkoff (TOV) equation~\cite{Tolman:1939jz,Oppenheimer:1939ne} for equilibrium configurations, we exploited Chandrasekhar’s adiabatic index criterion~\cite{Chandrasekhar:1964zza,Chandrasekhar:1964zzb} to find sufficient conditions for collapse. In the classical regime, the results indicate that the core temperature must be at least $\sim 10\%$ of the particle mass to trigger black hole formation. This condition implies that the formation of seed black holes as massive as $10^9{\rm\,M}_\odot$ requires particle mass above the ${\rm keV}$ scale.

In this work, we extend our previous study~\cite{Feng:2021rst} to the quantum regime and investigate the conditions for dynamical instability when Fermi degeneracy pressure becomes important. We model the system using a truncated Fermi–Dirac distribution and solve the TOV equation to obtain equilibrium configurations for a self-gravitating sphere, assessing their stability following Chandrasekhar’s criterion. Unlike in the classical regime, where thermal pressure drives the radial instability of a gravothermal system, in the quantum regime Fermi pressure can instead dominate and trigger the instability. Consequently, the boundary temperature required for collapse is reduced. In this limit, the critical mass becomes insensitive to the degree of degeneracy and depends solely on the particle mass. We discuss the implications for the formation of massive black holes in the early Universe through the collapse of dark matter.

The rest of the paper is organized as follows. In Sec.~\ref{sec:method}, we present our model, including the truncated Fermi–Dirac distribution, the TOV equation, and the equation of state. In Sec.~\ref{sec:fermi}, we discuss the instability conditions in the quantum, classical, and mixed regimes. In Sec.~\ref{sec:instablity}, we present numerical results based on solutions to the TOV equation and map out the parameter space of stable and unstable configurations. In Sec.~\ref{sec:mass}, we determine the critical mass and discuss its implications for the formation of massive black holes through dark matter collapse in the early Universe. We conclude in Sec.~\ref{sec:con}. Appendix~\ref{sec:gr} derives gravitational and thermodynamical constraints on the gravothermal system in general relativity. Appendix~\ref{sec:index} provides expressions for the averaged adiabatic index and the critical index. Appendix~\ref{sec:solution} lists the numerical results in tabulated form.

\section{The Fermi--Dirac Model}
\label{sec:method}

We introduce a truncated Fermi--Dirac distribution function for the ideal-gas sphere with a boundary radius of $R$~\cite{Ruffini:1983,Feng:2021rst},
\begin{equation}
\label{eq:distribution}
f(\epsilon)=
\begin{cases}
\displaystyle \frac{1-e^{(\epsilon-\epsilon_c)/k_B T}}{e^{(\epsilon-\mu)/k_B T}+1}, & \epsilon \le \epsilon_c,\\[4pt]
0, & \text{otherwise.}
\end{cases}
\end{equation}
where $\epsilon$ is the kinetic energy, $\epsilon_c$ the cutoff energy, $T$ the temperature, and $\mu$ the chemical potential with the rest energy $mc^2$ being subtracted. These physical quantities depend on the radius $r$ for $r\leq R$. $k_B$ is the Boltzmann constant. The quantum effect becomes important when $e^{(\epsilon-\mu)/k_BT}\lesssim1$, or $\mu/k_BT\gtrsim0$ as $\epsilon\geq0$.

The presence of the gravitational potential of the self-gravitating sphere is encoded in the radial dependence of the cutoff energy $\epsilon_c(r)$. Additionally, the temperature of the sphere follow a radial profile as~\cite{Feng:2021rst}
\begin{equation}
\label{eq:temp}
T(r)=\left[1+\frac{\epsilon_c(r)}{mc^2}\right]T(R),
\end{equation}
where $T(R)$ is the temperature at the outer boundary $R$ of the sphere. As a self-gravitating system, the sphere does not follow a global isothermal distribution due to the effect of gravitational redshifts~\cite{Tolman:1930ona}; see also Appendix~\ref{sec:gr} for derivation.   

For convenience, we introduce the dimensionless quantities~\cite{Feng:2021rst},
\begin{eqnarray}
x(r)&\equiv&\epsilon(r)/{k_B T(r)},\\
w(r)&\equiv&{\epsilon_c(r)}/{k_B T(r)},\\ 
\alpha(r)&\equiv&{\mu(r)}/{k_BT(r)},\\
b&\equiv& {k_B T(R)}/{mc^2}.
\end{eqnarray}

For the gravothermal system we consider, the cutoff energy at radius $r$, for which $r\leq R$, is related to the degeneracy difference through the Tolman--Klein law~\cite{Tolman:1930ona,Klein:1949}.
\begin{equation}
\label{eq:cutoff}
w(r)=\alpha(r)-\alpha (R).
\end{equation}
In the limit $r=R$, we see that $w(R)=0$, i.e., the cutoff energy vanishes at the outer boundary as expected for a self-gravitating system. Consider $\alpha(R)=0$, such that particles in the sphere are completely degenerate $\alpha(r)>0$, the cutoff energy and the degeneracy take the same value as indicated in Eq.~\ref{eq:cutoff}, and $\epsilon_c(r)=\mu(r)>0$ is the energy required to move a particle from the outer boundary $R$ to the radius $r$. For the model we consider, we take a specific form of the truncation as in Eq.~\ref{eq:distribution}, but the relations shown in Eqs.~\ref{eq:temp} and~\ref{eq:cutoff} are generic as they are based on laws of thermodynamics and general relativity.

We can obtain the profile of $w(r)$ by solving the TOV equation~\cite{Feng:2021rst}
\begin{equation}
\label{eq:tov}
\frac{{\rm d} w(r)}{{\rm d} r}=-\f{G}{r c^2}\left[\frac{1-bw(r)}{b}\right]\frac{4\pi p(r)r^3+{M(r)c^2}}{rc^2-2 GM(r)},
\end{equation}
where the enclosed mass is $M(r)=4\pi\int_0^r r'^2\rho{\rm~d}r'$ and $p(r)$ is the pressure profile. The associated boundary conditions are: $M(0)=0$ and $w(0)=\epsilon_c(0)/k_B T(0)$ at $r=0$; $M(R)=M$ and $w(R)=0$ at $r=R$. 

The model has three parameters, namely, the boundary temperature $b=k_B T(R)/mc^2$ and degeneracy (chemical potential) $\alpha(R)=\mu(R)/k_BT(R)$ at $r=R$, as well as the central cutoff energy $w(0)=\epsilon_c(0)/k_B T(0)$ at $r=0$. For a given set of these parameters, we solve the TOV equation Eq.~\ref{eq:tov} to obtain the $w(r)$ profile. Then, we use the distribution in Eq.~\ref{eq:distribution} to determine radial profiles for the number density $n(r)$, energy density $\rho(r)c^2$, internal energy density $u(r)$, and pressure $p(r)$ as
\begin{widetext}
\begin{equation}
\label{eq:eos}
\begin{aligned}
n(r)=&4\sqrt{2}\pi gm^3(c^3/h^3)e^{\alpha (R)}\left(\frac{b}{1-bw}\right)^{3/2}\int^{w}_{0}\frac{e^{w-x}-1}{1+ e^{\alpha(R)+w-x}}\left(1+\frac{bx/2}{1-bw}\right)^{1/2}\left(1+\frac{bx}{1-bw}\right)x^{1/2}{\rm d} x,\\
\rho(r)=&4\sqrt{2}\pi gm^4(c^3/h^3)e^{\alpha (R)}\left(\frac{b}{1-bw}\right)^{3/2}\int^{w}_{0}\frac{e^{w-x}-1}{1+ e^{\alpha(R)+w-x}}\left(1+\frac{bx/2}{1-bw}\right)^{1/2}\left(1+\frac{bx}{1-bw}\right)^2x^{1/2}{\rm d} x,\\
u(r)=&4\sqrt{2}\pi gm^4(c^5/h^3)e^{\alpha (R)}\left(\frac{b}{1-bw}\right)^{5/2}\int^{w}_{0}\frac{e^{w-x}-1}{1+ e^{\alpha(R)+w-x}}\left(1+\frac{bx/2}{1-bw}\right)^{1/2}\left(1+\frac{bx}{1-bw}\right)x^{3/2}{\rm d} x,\\
p(r)=&(8/3)\sqrt{2}\pi gm^4(c^5/h^3)e^{\alpha (R)}\left(\frac{b}{1-bw}\right)^{5/2}\int^{w}_{0}\frac{e^{w-x}-1}{1+ e^{\alpha(R)+w-x}}\left(1+\frac{bx/2}{1-bw}\right)^{3/2}x^{3/2}{\rm d} x,
\end{aligned}
\end{equation}
\end{widetext}
respectively, where $g=2s+1$ is the spin multiplicity of the gas particle, and $h$ is the Planck constant. In addition, the characteristic size of the system can be determined by the fiducial length in numerical implementation,
\begin{equation}
\label{eq:Length_scale}
\zeta=\lambda_C \left(\frac{m_{\text{Pl}}}{m}\right)\left(\frac{8\pi^3}{ge^{\alpha (R)}}\right)^{1/2}
~{\rm with}\quad
r=\zeta\hat{r},
\end{equation}
where $m_{\text{Pl}}=(\hbar c/G)^{1/2}$ is the Planck mass, $\lambda_C=\hbar/mc$ is the reduced Compton wavelength of the particle ($\hbar=h/2\pi$), and $\hat{r}$ is the dimensionless radius. Furthermore, we can express the dimensional quantities of the system using their corresponding dimensionless counterpart denoted with a ``hat,'' {i.e.}, $n=(c^2/Gm\zeta^2)\hat{n}$, $\rho=(c^2/G\zeta^2)\hat{\rho}$, $u=(c^4/G\zeta^2)\hat{u}$, $p=(c^4/G\zeta^2)\hat{p}$, and $M=(c^2\zeta/G)\hat{M}$.

%%%
\begin{figure}[h]
\centering
   \includegraphics[width=0.48\textwidth]{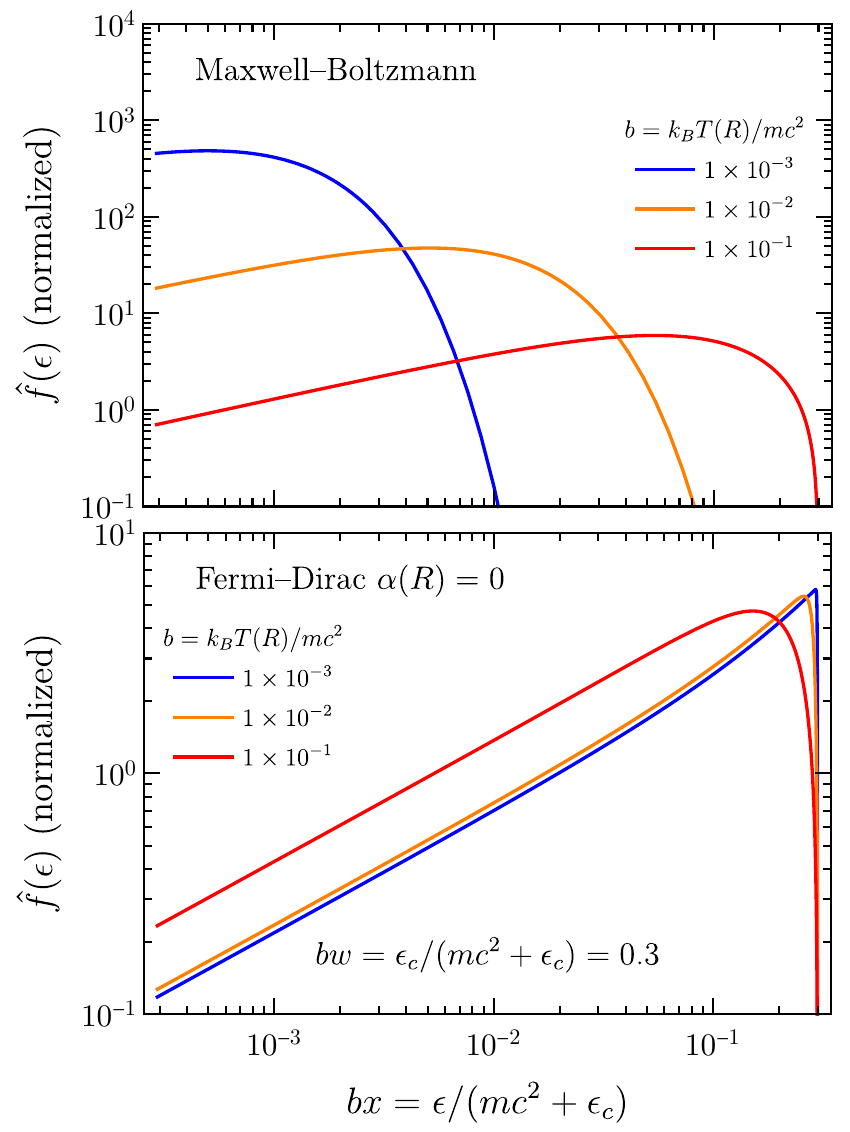}
   \caption{Normalized classical Maxwell–Boltzmann distribution (top) and quantum Fermi--Dirac distribution with $\alpha(R)=0$ (bottom). The cutoff energy is fixed at $bw= 0.3$, and the boundary temperature varies as $b = k_B T(R) / mc^2 = 10^{-3}$ (blue), $10^{-2}$ (orange), and $10^{-1}$ (red). In the classical limit, the energy peak shifts toward higher energies as the temperature increases, whereas in the quantum limit, the trend reverses as the fermions become fully degenerate.}
   \label{fig:dist}
\end{figure}

\section{Fermi Degeneracy Pressure}
\label{sec:fermi}

Before presenting our full numerical results, it is useful to first discuss the dynamical instability that arises when Fermi pressure becomes significant. We consider the thermodynamic quantities in Eq.~\ref{eq:eos} at a fixed radius, without requiring solutions to the TOV equation. Since the condition in Eq.~\ref{eq:Norm_cutoff} holds for any radius as long as $r<R$, we will omit the explicit ``$r$" dependence in this Section. In Sec.~\ref{sec:instablity}, we will extend the analysis to the entire sphere by taking volume averages after solving the TOV equation.

To trigger the instability, the system needs to be in the relativistic limit, which imposes conditions on the boundary temperature $b$ and the cutoff energy $w$; both of them are free parameters in the model. Consider their product~\cite{Feng:2021rst}
\begin{equation}
\label{eq:Norm_cutoff}
bw=\frac{\epsilon_c/mc^2}{1+\epsilon_c/mc^2}.
\end{equation}
In the ultrarelativistic limit $\epsilon_c\gg mc^2$, $bw\rightarrow 1$; in the non-relativistic limit,  $bw\rightarrow 0$. Since $b= k_B T(R)/mc^2$ determines the temperature at the boundary, a high $b$ value indicates a hot thermal bath. For a small cutoff energy, a high boundary temperature would be needed to drive the sphere into the relativistic regime $bw\approx0.1\textup{--}1$, or vice versa.

To illustrate the role of Fermi pressure, we first examine the truncated Maxwell–Boltzmann~\cite{Feng:2021rst} and Fermi–Dirac distributions. For the latter, we set $\alpha(R) = \mu(R)/k_B T(R) = 0$, corresponding to a completely degenerate system. We fix $bw = 0.3$ and vary $b$ as $10^{-3}$, $10^{-2}$, and $10^{-1}$. Using the number density profile given in Eq.~\ref{eq:eos}, we then compute the normalized distribution function as 
\begin{equation}
\label{eq:norm}
\hat{f}(\epsilon)=\hat{f}(bx)=
\frac{1}{n}\frac{{\rm\,d}n}{b{\rm\,d}x}
\end{equation}
as a function of $bx=\epsilon/(mc^2+\epsilon_c)$.  

In Fig.~\ref{fig:dist}, we compare the normalized distribution function in the classical (top) and quantum (bottom) limits in the relativistic regime $bw=0.3$ for the boundary temperature $b=10^{-3}$ (blue), $10^{-2}$ (orange), and $10^{-1}$ (red). In the classical Maxwell-Boltzmann limit, most of the particles are non-relativistic with $bx\lesssim10^{-2}$ for $b=10^{-3}$. As $b$ increases to $0.1$, more and more particles occupy the high energy state. Thus to trigger the instability in this case, one must increase the boundary temperature $b$ such that the majority of the particles can exist in the high energy state $bx\sim0.1$. On the other hand, for the quantum limit $\alpha(R)=0$, the distribution is peaked at $bx\approx0.3$ even at the low temperature $b=10^{-3}$. As $b$ increases further, the distribution changes mildly as in this limit the Pauli blocking effect dominates.

%%%
\begin{figure*}[t]
\centering
   \includegraphics[width=\textwidth]{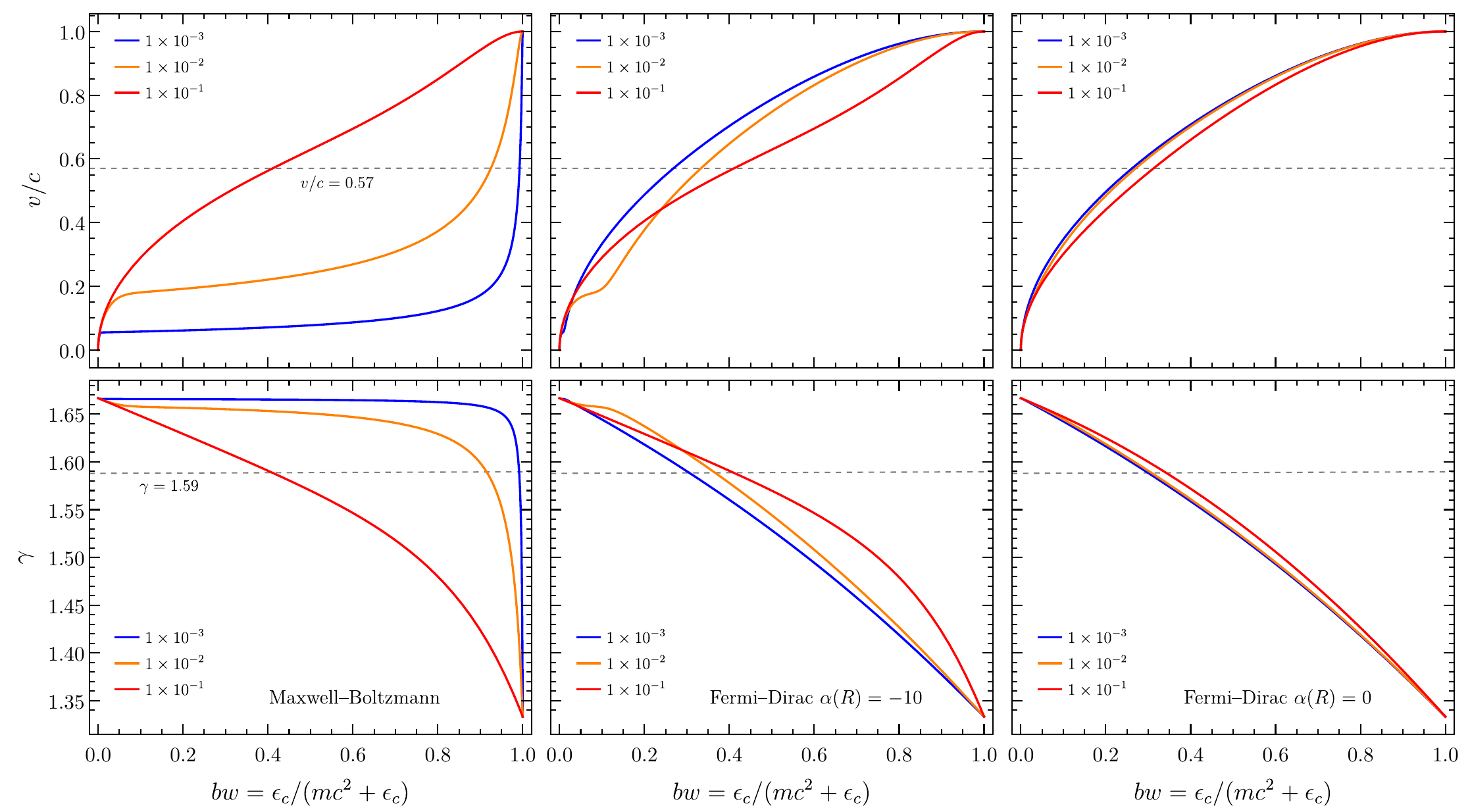}
   \caption{Three-dimensional velocity dispersion $v/c$ (top) and adiabatic index $\gamma$ (bottom) as functions of the cutoff energy $bw$ for the Maxwell–Boltzmann distribution (left) and Fermi--Dirac distribution with $\alpha(R)=-10$ (middle) and $0$ (right), spanning the classical to quantum regimes. Results are shown for boundary temperatures $b = k_B T(R)/mc^2 = 10^{-3}$ (blue), $10^{-2}$ (orange), and $10^{-1}$ (red). The horizontal dashed lines in the top and bottom panels indicate $v/c = 0.57$ and $\gamma = 1.59$, respectively. Dynamical instability occurs when $v/c \gtrsim 0.57$ and $\gamma \lesssim 1.59$~\cite{Feng:2021rst}.
   }
   \label{fig:eosFD}
\end{figure*}
%%%

Furthermore, we evaluate the velocity dispersion $v/c=\sqrt{3p/\rho c^2}$ and adiabatic index $\gamma=1+p/u$, where pressure $p$, energy density $\rho c^2$, and internal energy $u$ are given in Eq.~\ref{eq:eos}. We consider the Maxwell--Boltzmann distribution and the Fermi--Dirac distribution with $\alpha(R)=\mu(R)/k_BT(R)=-10$ and $0$, ranging from the classical to quantum limits. The $b$ values are $b=k_BT(R)/mc^2=10^{-3}$, $10^{-2}$, and $10^{-1}$, the same as those in Fig.~\ref{fig:dist}. 

In Fig.~\ref{fig:eosFD}, we show the velocity dispersion (top) and adiabatic index (bottom) as functions of the cutoff energy. In the classical Maxwell–Boltzmann limit, reaching the instability threshold ($v/c \gtrsim 0.57$ and $\gamma \lesssim 1.59$) requires substantially increasing either the boundary temperature or the cutoff energy, or both. For a given cutoff energy $bw$, the equation of state is primarily governed by the boundary temperature $b$: higher temperatures lead to a softer state, characterized by larger $v/c$ and smaller $\gamma$. At low temperatures ($b=10^{-3}$), the state can only be softened if the cutoff energy $bw$ approaches unity. As the degeneracy parameter $\alpha(R)$ increases toward $0$, corresponding to a fully degenerate configuration, the temperature dependence of the equation of state weakens because the Fermi pressure becomes dominant. In the quantum limit, the instability condition requires $bw \gtrsim 0.3$ for the low-temperature case ($b=10^{-3}$), as the equation of state softens further due to Fermi pressure. This softening is crucial for triggering instability and initiating collapse at low temperatures.

\section{Dynamical Instability}

\label{sec:instablity}

\begin{figure*}[t]
\centering
   \includegraphics[width=0.95\textwidth]{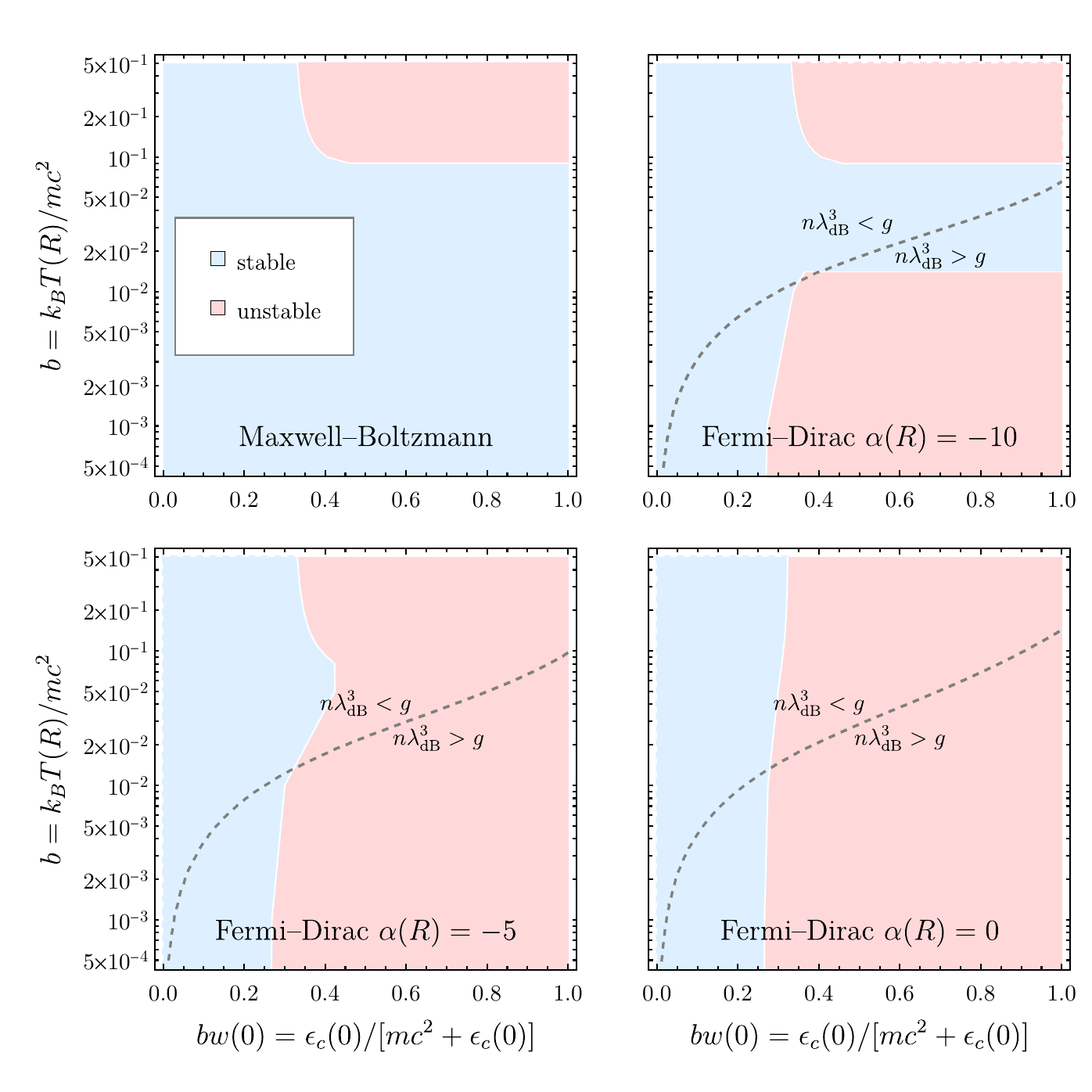}
   \caption{Parameter space for stable ($\langle \gamma \rangle > \gamma_{\rm cr}$, blue shaded) and unstable ($\langle \gamma \rangle < \gamma_{\rm cr}$, red shaded) configurations in the plane of central cutoff energy $bw(0)$ versus boundary temperature $b$ for the Maxwell–Boltzmann distribution and the Fermi--Dirac distribution with $\alpha(R)=-10$, $-5$ and $0$, spanning the transition from classical to quantum regimes. In the classical regime, instability occurs only when $b \gtrsim 0.1$. When Fermi pressure becomes important, instability can also arise in the low-temperature regime $b \lesssim 10^{-2}$. A gap remains between the low- and high-temperature instability regions until $\alpha(R) \gtrsim -5$. The gray dashed line marks the boundary between classical ($n\lambda_{\rm dB}^3<g$) and quantum ($n\lambda_{\rm dB}^3>g$) regimes based on wave function overlap.}
   \label{fig:regionsFD}
\end{figure*}
%%%

In this Section, we present results from our full numerical study by solving the TOV equation. For a given set of the $\{b, w(0), \alpha(R)\}$ values, we follow the Chandrasekhar criterion~\cite{Chandrasekhar:1964zza} as in~\cite{Feng:2021rst} and find the marginally stable configuration satisfying $\langle\gamma\rangle=\gamma_{\rm cr}$, where $\langle\gamma\rangle$ is the pressure-averaged adiabatic index with $\gamma=1+p/u$ and $\gamma_{\rm cr}$ the corresponding critical value; see Appendix~\ref{sec:index} for their detailed expressions. The instability is triggered when $\langle\gamma\rangle<\gamma_{\rm cr}$. In practice, we vary the model parameters in the following ranges: the boundary temperature $b=(10^{-5}, 5)$, the degeneracy $\alpha(R)=(-50,1)$. To find the corresponding central cutoff $w(0)$ that triggers the instability, we use the fourth-order Runge-Kutta algorithm to solve the TOV equation Eq.~\ref{eq:tov} and obtain the thermodynamical quantities shown in Eq.~\ref{eq:eos}. We then identify marginally stable configurations that satisfy $\langle\gamma\rangle=\gamma_{\rm cr}$. The numerical results are tabulated in Appendix~\ref{sec:solution}. We have also checked that marginally stable configurations are indistinguishable for a Maxwell--Boltzmann distribution and a Fermi--Dirac distribution with $\alpha(R)=-50$. In what follows, we highlight main findings and discuss their implications.

In Fig.~\ref{fig:regionsFD}, we show the stable and unstable regions in the plane of the boundary temperature $b$ and the central cutoff $bw(0)$ for the degeneracy $\alpha(R)=-50$, $-10$, $-5$, and $0$, covering the classical to quantum limits. In the classical limit ($\alpha(R)=-50$), the instability can only occur when the boundary temperature is high $b=k_B T(R)/mc^2>9\times10^{-2}$ and the central cutoff is $bw(0)\gtrsim0.35$ (relativistic regime, see Eq.~\ref{eq:Norm_cutoff}). As the degeneracy $\alpha(R)$ increases to $-10$, another unstable region appears for $b<1.5\times10^{-2}$ due to Fermi pressure, while the requirement on the central cutoff remains almost the same. In this case, the unstable region with high boundary temperature is still preserved, where the thermal pressure is the dominant source of the instability. Interestingly, there is a gap in the boundary temperature between the two unstable regions. In the gap region, both thermal pressure and Fermi pressure are relevant, but overall they are not strong enough to trigger the instability. As $\alpha(R)$ further increases, the gap shrinks and vanishes for $\alpha(R)\gtrsim-5$. For $\alpha(R)=0$, both regions join smoothly and the instability condition becomes insensitive to the boundary temperature $b$, as long as the central cutoff $bw(0)\gtrsim0.3$.

Additionally, we check the border between classical and quantum regimes set by the condition $n\lambda^3_{\rm dB}=g$, where $n$ is the number density of the particles, $\lambda_{\rm dB}$ is the de Broglie thermal wavelength, and $g=2s+1$ is the spin multiplicity. For the quamtum effect to be important, the wavelength should be larger than the average distance between particles up to the $g$ factor, i.e., $n\lambda^3_{\rm dB}>g$. The de Broglie thermal wavelength is calculated as~\cite{Feng:2021rst}. 
\begin{eqnarray}
\label{eq:labmdadb}
\lambda_{\rm dB}(r)&&= \lambda_{\rm C} \left[\left(1+\frac{3k_B T(r)}{2 mc^2}\right)^2-1\right]^{-1/2} \\
&&=\lambda_{\rm C}\left[\left(1+\frac{3b/2}{1-bw(r)}\right)^2-1\right]^{-1/2},
\end{eqnarray}
where $\lambda_{\rm C}=\hbar/mc$ is the reduced Compton wavelength.
Using Eq.~\ref{eq:Length_scale}, we have the following relation
\begin{equation}
\label{eq:nlambdac}
\frac{n(r)\lambda^3_{\rm C}}{g}=\frac{\hat{n}(b,w(r),\alpha(R))}{8\pi^3}e^{\alpha(R)}.
\end{equation}
By Eqs.~\ref{eq:labmdadb} and ~\ref{eq:nlambdac}, we find the border between the classical and quantum regimes $n(0)\lambda^3_{\rm dB}(0)=g$ in the $b\textup{--}bw(0)$ plane; shown in Fig.~\ref{fig:regionsFD} as dashed curves in the last three panels. In the high-temperature regime where $n(0)\lambda_{\rm dB}^3 (0) < g$, the instability is primarily driven by thermal pressure, whereas in the opposite regime it is dominated by Fermi pressure.

Lastly, we comment on the compactness. For $\alpha(R) = -10$ and $-5$, the compactness for the marginally stable configurations $C = \hat{M}/{\hat R}$ increases with temperature in the high-temperature regime $0.1 \lesssim b \lesssim 0.5$, ranging from $\simeq 0.04$ to $0.075$. In the low-temperature regime $b \lesssim 10^{-2}$, the core becomes partially degenerate, and $C$ asymptotes to $\simeq 0.128$, indicating a more compact configuration due to Fermi pressure. In contrast, for completely degenerate cores with $\alpha(R) = 0$, the critical compactness decreases with temperature across all regimes; see Appendix~\ref{sec:solution} for details.

\section{Black Hole Mass} 
\label{sec:mass}

With our numerical results as tabulated in Appendix~\ref{sec:solution}, we can calculate the critical mass for marginally stable configurations satisfying $\left<\gamma\right>=\gamma_{\rm cr}$. 
Since the dynamical evolution after reaching the instability leads to horizon formation~\cite{Saijo:2002qt,Shibata:2002br,Shapiro:2002kk}, we assume here that a black hole with the critical mass will form following the onset of instability.
From Eq.~\ref{eq:Length_scale}, the gravitational mass of the bound sphere is given by
\begin{equation}
\label{eq:mass}
M=\frac{m_{\rm Pl}^3}{m^2}\left(\frac{8\pi^3}{ge^{\alpha(R)}}\right)^{1/2}\hat{M}. 
\end{equation}
For given values of degeneracy $\alpha(R)$ and temperature $b=k_BT(R)/mc^2$ at the boundary, there is a corresponding value of $\hat {M}$ for the marginally stable configuration; see Appendix~\ref{sec:solution} for details. Consider the high-temperature regime $0.1\leq b\leq0.5$, $\hat{M}\approx(1.42\textup{--}5.95)\times10^{-2}$ for $\alpha(R)=-50$, $\hat{M}\approx(1.42\textup{--}5.95)\times10^{-2}$ for $\alpha(R)=-10$, and $\hat{M}\approx(4.64\textup{--}9.35)\times10^{-2}$ for $\alpha(R)=0$. Therefore, in the high-temperature regime, the critical mass is mainly controlled by $\alpha (R)\leq0$, up to a unity factor,
\begin{equation}
\label{eq:critical1}
M\gtrsim\frac{0.2}{\sqrt{g}} \frac{m^3_{\rm Pl}}{m^2}e^{-\alpha(R)/2}\approx2.3\times10^5\,{\rm M_\odot}\left(\frac{1\,{\rm MeV}}{mc^2}\right)^2e^{-\alpha(R)/2}.
\end{equation}

On the other hand, for the low-temperature regime $b\leq10^{-2}$, $\hat{M}/\sqrt{e^{\alpha(R)}}\approx0.034$, which is {\rm universal} for $\alpha(R)$ from $-10$ to $0$ as shown in Appendix~\ref{sec:solution}. From Eq.~\ref{eq:mass}, we find the critical mass as 
\begin{equation}
\label{eq:critical2}
M\gtrsim \frac{0.54}{\sqrt{g}}\frac{m^3_{\rm Pl}}{m^2}\approx5.7\times10^5\,{\rm M_\odot}\left(\frac{1\,{\rm MeV}}{mc^2}\right)^2. 
\end{equation}
When the degeneracy is large, Fermi pressure is the dominant source to trigger the instability even at a relatively low boundary temperature. In this regime, the critical mass is {\it independent} of the boundary temperature. In contrast, for the classical Maxwell--Boltzmann distribution, where thermal pressure is the only source of the instability and the only way to trigger the instability is to increase the boundary temperature. Note that for $\alpha(R)=-50$, the system is in the classical limit and follows the Maxwell--Boltzmann distribution, there are no marginally stable configurations in the low-temperature regime, as indicated in Fig.~\ref{fig:regionsFD}. We have checked that these features are generic and insensitive to the specific truncation model in Eq.~\ref{eq:distribution}.

%%%
\begin{figure*}[t]
\centering
   \includegraphics[width=1.0\textwidth]{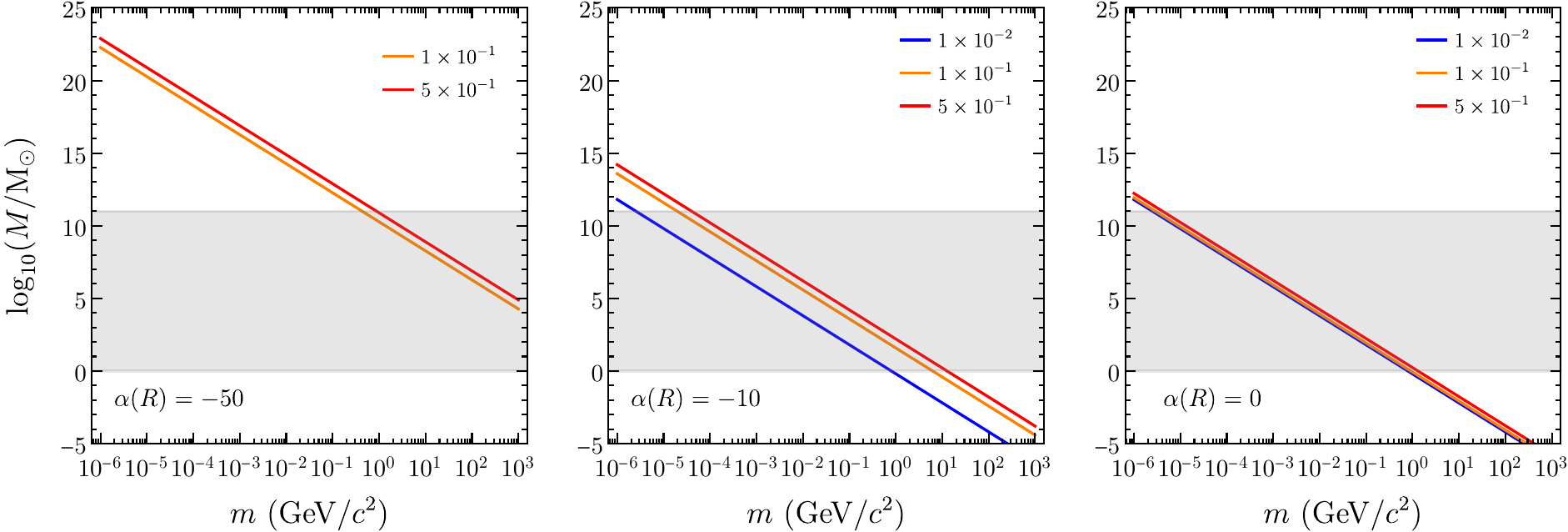}
   \caption{Critical mass $M$ as a function of particle mass $m$ for the degeneracy $\alpha(R)=-50$ (left), $-10$ (middle), and $0$ (right), obtained by varying the boundary temperature $b=0.5$ (red), $10^{-1}$ (orange), and $10^{-2}$ (blue). At the low temperature $b=10^{-2}$, marginally stable configurations are not found for $\alpha(R)=-50$, where the classical Maxwell–Boltzmann distribution is recovered, while the critical mass converges for $\alpha(R)=-10$ and $0$ as the core becomes completely degenerate. The shaded gray band in each panel indicates the black hole mass range of astrophysical interest. The lower bound, $1\,{\rm M_\odot}$, corresponds to stellar-mass black holes, while the upper bound, $10^{11}{\rm\,M_\odot}$, approximates the most massive black holes discovered.}
   \label{fig:scales}
\end{figure*}
%%%

Fig.~\ref{fig:scales} shows the critical mass as a function of particle mass for $\alpha(R)=-50$ (left), $-10$ (middle), and $0$ (right) for $b=0.5$ (red), $0.1$ (orange), and $10^{-2}$ (blue). For a given $\alpha(R)$, the critical mass increases with the boundary temperature $b$. For $\alpha(R)=-50$, the instability cannot be triggered at a low temperature $b=10^{-2}$, but this case has the highest critical mass at the high-temperature regime among the three examples. Thus in the classical regime, black hole seeds tend to be more massive. In the low-temperature regime $b=10^{-2}$, the critical mass converges and becomes insensitive to $\alpha(R)$. In this completely degenerate regime, the critical mass reaches the minimum for a given particle mass.

As an application, we use Eq.~\ref{eq:critical2} to estimate the mass of marginally stable neutron stars as $0.7\,{\rm M_\odot}$, by taking $m \approx 0.94\,{\rm GeV}$, assuming the neutrons form a degenerate, cold Fermi gas. The corresponding compactness is $C \approx 0.12$ (Table~\ref{table:FD0}). This estimate agrees remarkably well with the results of Ref.~\cite{Oppenheimer:1939ne}. Observationally, neutron star masses are typically around $2\,{\rm M_\odot}$ with compactness $C \approx 0.25$. The discrepancy arises because the ideal-gas assumption breaks down in the stellar core due to strong nuclear interactions. The repulsive nuclear forces stiffen the equation of state, leading to larger masses and higher compactness.

Now we consider scenarios where dark matter clumps collapse into black holes due to gravothermal evolution, such as in the self-interacting dark matter framework~\cite{Balberg:2001qg,Feng:2020kxv,Feng:2021rst}. For fermonic dark matter, there is a lower bound on the particle mass, i.e., the Tremaine-Gunn bound~\cite{Tremaine:1979we}, below which dark matter halos cannot form. Recent studies found that the lower bound is $m\gtrsim0.4\,{\rm~keV}/c^2$~\cite{Boyarsky:2008ju}, although it could be relaxed if many distinct species of fermions are allowed~\cite{Davoudiasl:2020uig}. Taking $m=0.4\,{\rm keV/c^2}$ in Eq.~\ref{eq:critical2}, we find that the critical mass is $3.9\times10^{12}\,{\rm M_\odot}$. This would exceed the mass of the most massive supermassive black holes observed to date, $\sim10^{11}{\rm\,M_\odot}$~\cite{Brockamp:2016aa,Dullo:2017aa}. For a thermally produced dark matter particle, the gravitational lensing constraint from JWST data sets a lower limit on its mass of $6.1\,{\rm keV}/c^2$~\cite{Keeley:2024brx}, corresponding to a critical mass of $1.67\times10^{10}\,{\rm M_\odot}$ from Eq.~\ref{eq:critical2}. Conversely, if $\sim10^7\,{\rm M_\odot}$ black holes associated with the JWST little red dots~\cite{Matthee:2023utn,Harikane:2023aa,Kocevski:2023aa,Pacucci:2023oci,Maiolino:2023bpi} are formed through the collapse of degenerate fermionic cores, the particle mass is estimated as $m\sim100{\rm\,keV/c^2}$~\cite{Arguelles:2023hab,Arguelles:2023kqw}. 

Our model relies on the ideal-gas assumption, where the interactions among particles do not modify the equation of state. In a typical self-interacting dark matter scenario, there is a light mediator and its mass $m_\phi$ is smaller than dark matter particle mass $m$. Observations spanning dwarf galaxies to galaxy clusters require the self-interaction cross section to be velocity-dependent, ranging from $\sim 10\text{--}100\,{\rm cm^2/g}$ at $v \sim 10\,{\rm km/s}$ to $\sim 0.1\,{\rm cm^2/g}$ at $v \sim 1000\,{\rm km/s}$, which implies $m/m_\phi \sim 1000$~\cite{Tulin:2017ara}. For the ideal-gas assumption to hold, the average distance between particles in the core should be larger than the interaction range set by $\hbar/m_\phi c$. The average number density of particles within a sphere is $3M/(4\pi R^3 m)$. Hence the average distance $d$
\begin{eqnarray}
d&&\approx\left(\frac{4\pi}{3}\frac{m}{M}R^3\right)^{1/3}\notag\\
&&=\left(\frac{4\pi}{3}\right)^{1/3}\frac{\hbar}{mc}\left(\frac{8\pi^3}{ge^{\alpha(R)}}\right)^{1/3}\frac{\hat{R}}{\hat{M}^{1/3}},
\end{eqnarray}
where we have used Eqs.~\ref{eq:Length_scale} and \ref{eq:mass}. By requiring $d>\hbar/m_\phi c$, we have 
\begin{equation}
\label{eq:mphi}
\left(\frac{4\pi}{3}\right)^{1/3}\left(\frac{8\pi^3}{ge^{\alpha(R)}}\right)^{1/3}\frac{\hat{R}}{\hat{M}^{1/3}}\gtrsim\frac{m}{m_\phi}. 
\end{equation}
For marginally stable configurations, $\hat{R}/\hat{M}^{1/3}$ depends on $\alpha(R)$ and $b$ as shown in Appendix~\ref{sec:solution}. Consider $\alpha(R)=-10$, we have $\hat{R}/\hat{M}^{1/3}\sim2$ at high temperatures $0.1\leq b\leq0.5$ and obtain $\alpha(R)\lesssim-10$ from Eq.~\ref{eq:mphi}. In the low-temperature regime $b\leq10^{-2}$, $\hat{R}/\hat{M}^{1/3}$ becomes smaller, $\alpha(R)$ needs to be smaller for the condition in Eq.~\ref{eq:mphi} to be satisfied. Therefore, for viable self-interacting dark matter scenarios, particle interactions with a light mediator may not be negligible in the completely degenerate limit $\alpha(R)\rightarrow0$ for examining the instability and deriving the critical mass. 

Ref.~\cite{Gresham:2018rqo} applied the method of mean field theory in the limit of zero temperature and derived the critical mass with such interactions, in which they exploited the turning-point method to assess the instability condition. We note that our critical mass, Eq.~\ref{eq:critical2}, by Chandrasekhar's criterion in the low-temperature regime agrees well with their cases without interactions $M\approx0.384m_{\rm Pl}^3/m^2$. It would be interesting to extend their analysis to high temperatures with the method laid out in this work.

\section{Conclusions}
\label{sec:con}

In this work, we investigated dynamical instability in the quantum regime, where Fermi pressure becomes significant. We modeled the system using a truncated Fermi–Dirac distribution and examined the instability conditions based on Chandrasekhar’s criterion. In the classical limit, the instability is driven by thermal pressure, requiring a high boundary temperature for the sphere to reach the instability threshold. In contrast, in the quantum regime, Fermi pressure can become the dominant factor triggering the onset of instability, allowing the boundary temperature to be lower. This opens a new parameter space for exploring the formation of black holes. We numerically solved the TOV equation over a wide range of model parameters, including classical, mixed, and quantum regimes, and identified the critical configurations that are marginally stable. The numerical results are publicly available.

We also examined the critical mass required for the sphere to collapse into a black hole. In the classical and mixed regimes, a high boundary temperature is favored for the onset of instability, and the critical mass increases with boundary temperature while being inversely proportional to the square of the particle mass. In contrast, in the quantum limit, where Fermi pressure is the dominant driver of instability, the critical mass becomes insensitive to the boundary temperature and depends solely on the particle mass. The critical mass derived in the quantum limit thus represents the lower bound on the black hole mass for a given particle mass. We further discussed the implications of these results for black hole formation in scenarios where fermionic dark matter collapses into black holes. 

In future work, this study can be extended to include interactions among fermionic particles. As in the case of neutron stars, such interactions would modify the equation of state and alter the instability conditions derived under the assumption of an ideal-gas model. In particular, it would be interesting to investigate how dark forces in self-interacting dark matter models influence these instability conditions.

\section*{Acknowledgments}

WXF was supported by Tsinghua’s Shuimu Scholar Fellowship and the China Postdoctoral Science Foundation under Grant No.\,2024M761594. HBY was supported by the U.S. Department of Energy under Grant No.\,DE-SC0008541 and the John Templeton Foundation under Grant No.\,63599. YZ was supported by the GRF Grants No.\,11302824 and No.\,11310925 from the Research Grants Council, University Grants Committee, and the Grants No.\,9610645 and No.\,7020130 from the City University of Hong Kong. The opinions expressed in this publication are those of the authors and do not necessarily reflect the views of the funding agencies.

\bibliographystyle{JHEP} 
\bibliography{diqc}

\onecolumngrid
\newpage
\appendix

\section{Gravitational and thermodynamical constraints}
\label{sec:gr}
Consider a perfect fluid $T_{\mu\nu}=(\rho c^2+p)u_\mu u_\nu+pg_{\mu\nu}$ in a spherically symmetric and static spacetime
\begin{equation}
{\rm d}s^2=g_{\alpha\beta}{\rm d}x^\alpha{\rm d}x^\beta
=-e^{2\Phi(r)}c^2{\rm d}t^2+e^{2\Lambda(r)}{\rm d}r^2+r^2({\rm d}\theta^2+\sin^2\theta~{\rm d}\phi^2).
\end{equation}
Solving the Einstein equation leads to the Tolman--Oppenheimer--Volkoff (TOV) equation
\begin{equation}\label{TOV1}
\frac{{\rm d}p}{{\rm d}r}=-(\rho c^2+p)\frac{{\rm d}\Phi}{{\rm d}r}=-(\rho c^2+p)\frac{G\left(4\pi pr^3+M(r)c^2\right)}{rc^2(rc^2-2GM(r))},
\end{equation}
with $M(r)=4\pi\int_0^r r'^2\rho(r'){\rm~d}r'$, from which one can obtain the solution
\begin{equation}\label{metric_soln}
e^{2\Phi(r)}=\exp\left(2\int^\infty_r\frac{G\left(4\pi p (r')r'^3+M(r')c^2\right)}{r'c^2(r'c^2-2GM(r'))}{\rm d}r'\right),~e^{2\Lambda(r)}=\left(1-\frac{2GM(r)}{rc^2}\right)^{-1}.
\end{equation}

The time-like Killing vector implies the ``energy'' of a particle of mass $m$ moving along a geodesic with three-speed $v$ in this spacetime is
\[
-mcu_{t}=mc^2(1-v^2/c^2)^{-1/2}e^{\Phi}\equiv(\epsilon+mc^2)e^{\Phi}=\text{constant},
\]
where $\epsilon$ is the kinetic energy.
Thus the cutoff energy (the kinetic energy required to escape to the boundary surface)
\begin{equation}\label{constraint_epsilon}
(\epsilon_c(r)+mc^2)e^{\Phi(r)}=mc^2e^{\Phi(R)},
\end{equation}
where we use $\epsilon_c(R)=0$ at $\Phi(r=R)$.

According to the first law of thermodynamics, ${\rm d}E=T{\rm d}S-p{\rm d}V+\theta{\rm d}N$ and the Euler theorem, $E=TS-pV+\theta N$, where $\theta=\mu+mc^2$ is the chemical potential including rest mass energy, we obtain the Gibbs--Duhem equation $S{\rm d}T-V{\rm d}p+N{\rm d}\theta=0$. By substituting the entropy $S=(E+pV-\theta N)/T$ into the Gibbs--Duhem equation, we obtain
\[
{\rm d}p=\left(\rho c^2+p\right)\frac{{\rm d}T}{T}+n\left({\rm d}\theta-\theta\frac{{\rm d}T}{T}\right),
\]
where $\rho c^2\equiv E/V$ and $n\equiv N/V$ are, respectively,  the local energy and number densities. Together with ${\rm d}p=-(\rho c^2+p){\rm d}\Phi$, we have
\[
n\left(d\theta-\theta\frac{{\rm d}T}{T}\right)+\left(\rho c^2+p\right)\left({\rm d}\Phi+\frac{{\rm d}T}{T}\right)=0.
\]
Since this equation must hold for any given $n$, $\rho$ and $p$, both terms in this equation must be identically zero. Hence,
\begin{equation}
\begin{cases}
(\mu+mc^2)e^{\Phi}=\text{constant}\quad(\text{Klein law}),\\
Te^{\Phi}=\text{constant}\quad(\text{Tolman law}).
\end{cases}
\end{equation}
In contrast to a Newtonian system, the isothermality condition under thermodynamic equilibrium can never be satisfied globally in a curved spacetime.

Thus we set
\begin{equation}\label{constraint_mu}
(\mu(r)+mc^2)e^{\Phi(r)}=(\mu(R)+mc^2)e^{\Phi(R)}
\end{equation}
and
\begin{equation}\label{constraint_T1}
T(r)e^{\Phi(r)}=T(R)e^{\Phi(R)}.
\end{equation}

For convenience, we now introduce the (dimensionless) boundary temperature and cutoff function
\begin{equation}
\begin{cases}
b=k_BT(R)/mc^2,\\
w(r)=\epsilon_c(r)/k_BT(r).
\end{cases}
\end{equation}
Combining Eq.~\ref{constraint_epsilon} and Eq.~\ref{constraint_T1}, we have 
\[
1-(\epsilon_c/mc^2)(T(R)/T)=1-bw=e^{\Phi-\Phi(R)},
\]
therefore $0\leq bw<1$ due to the positivity of $e^{\Phi}$. Furthermore, since the matching condition $\Phi(R)+\Lambda(R)=0$,
\[
(1-bw)^2=e^{2(\Phi-\Phi(R))}=e^{2\Phi}(1-2GM(R)/Rc^2)^{-1}.
\]
We obtain
\begin{equation}\label{constraint_w1} 
e^{2\Phi(r)}=(1-2GM(R)/Rc^2)(1-bw(r))^2,
\end{equation}
which relates $w(r)$ to the gravitational potential $\Phi(r)$ for $r\leq R$. Along with Eq.~\ref{constraint_T1}, we have
\begin{equation}\label{constraint_T2}
T(r)=\frac{T(R)}{1-bw(r)}=\left(1+\frac{\epsilon_c(r)}{mc^2}\right)T(R).
\end{equation}
Dividing Eq.~\ref{constraint_mu} by Eq.~\ref{constraint_T1}, we have
\[
\mu/k_BT-\mu(R)/k_BT(R)=mc^2/k_BT(R)-mc^2/k_BT=(1/b)-(1/b)(1-bw)=w.
\]
Hence the cutoff function is the degeneracy contrast to that of the boundary
\begin{equation}\label{constraint_w2}
w(r)=\alpha(r)-\alpha(R),
\end{equation}
where $\alpha=\mu/k_BT$.

%%%

\section{The adiabatic index}
\label{sec:index}
Given a system solved by Eqs.~\ref{TOV1} and \ref{metric_soln}, the corresponding critical adiabatic index is calculated by
\begin{align}
\gamma_{\text{cr}}\equiv&\frac{4}{3}+\frac{\int e^{3\Phi+\Lambda}[16p+(e^{2\Lambda}-1)(\rho+p)](e^{2\Lambda}-1)r^2{\rm d}r}{36\int e^{3\Phi+\Lambda}pr^2{\rm d}r}\nonumber\\
&+\frac{4\pi G\int e^{3(\Phi+\Lambda)}[8p+(e^{2\Lambda}+1)(\rho c^2+p)]pr^4{\rm d}r}{9c^4\int e^{3\Phi+\Lambda}pr^2{\rm d}r}+\frac{16\pi^2 G^2\int e^{3\Phi+5\Lambda}(\rho c^2+p)p^2r^6{\rm d}r}{9c^8\int e^{3\Phi+\Lambda}pr^2{\rm d}r}
\end{align}
and the pressure-averaged adiabatic index is 
\begin{equation}
\langle\gamma\rangle\equiv\frac{\int e^{3\Phi+\Lambda}\gamma(r) pr^2{\rm d}r}{\int e^{3\Phi+\Lambda}pr^2{\rm d}r}.
\end{equation}
The system is marginally stable when $\langle\gamma\rangle=\gamma_{\rm cr}$.

%%%
\section{Numerical results of the truncated Fermi--Dirac model}
\label{sec:solution}

In this Section, we present the numerical results for marginally stable configurations obtained from solutions of the TOV equation. The effects of pair production and annihilation of dark matter particles within the sphere are not included here. These processes may become relevant for $b > 0.1$, and we will address them in a forthcoming companion paper.

In Tables~\ref{table:MB}\textup{--}\ref{table:FDp1}, we show marginally stable configurations for $\alpha(R)$ ranging from $-50$ to $1$. These configurations satisfy the adiabatic index criterion $\langle\gamma\rangle=\gamma_{{\rm cr}}$. From the 1st to 12th columns, we show their boundary temperature $b=k_B T(R)/mc^2$, central cutoff function $w(0)=\epsilon_c(0)/k_BT(0)$, gravitational mass $\hat M$, system radius $\hat R$, compactness $C = GM(R)/c^2R=\hat{M}/\hat{R}$, central interior redshift $Z(0)=e^{-\Phi(0)}-1$, central energy cutoff $\epsilon_c(0)$, central energy density $\hat \rho(0)$, central pressure $\hat p(0)$, central velocity dispersion $v(0)$, boundary velocity dispersion $v(R)$, and $\langle\gamma\rangle=\gamma_{\text{cr}}$.

\begin{table}
  \caption{Marginally stable configurations at different boundary temperature $b=k_BT(R)/mc^2$ for a Fermi--Dirac distribution with $\alpha(R)=-50$, corresponding to a non-degenerate core. In this case, dynamical instability cannot be achieved for $b<0.09$. Marginally stable configurations are indistinguishable for a Fermi--Dirac distribution with $\alpha(R)=-50$ and a Maxwell--Boltzmann distribution for $b\geq0.09$.}
 \centering
 %\begin{ruledtabular}
 \resizebox{\columnwidth}{!}{
 \begin{tabular}{l l l l l l l l l l l l}
  \hline\hline
  % inserts table 
  %heading
  $b$  &  $w(0)$  & $\hat{M}$ & $\hat{R}$ & $C=\hat{M}/\hat{R}$ & $Z(0)$ & $\epsilon_c(0)/mc^2$ & $\hat{\rho}(0)$ & $\hat{p}(0)$ & $v(0)/c$ & $v(R)/c$ & $\langle\gamma\rangle=\gamma_{\text{cr}}$ \\ [0.5ex] 
  \hline\hline
$5.0$ & $6.48150\times10^{-2}$  & $2.06015\times10^{-1}$ & $2.59700\times10^{0}$ & $7.93280\times10^{-2}$ & $6.12906\times10^{-1}$ & $4.79448\times10^{-1}$ & $1.60344\times10^{-1}$ & $1.71013\times10^{-2}$ & $5.65650\times10^{-1}$ & $3.36500\times10^{-3}$ & 1.62358 \\
$3.0$ & $1.08185\times10^{-1}$  & $1.58573\times10^{-1}$ & $2.00600\times10^{0}$ & $7.90492\times10^{-2}$ & $6.13491\times10^{-1}$ & $4.80491\times10^{-1}$ & $2.71859\times10^{-1}$ & $2.89847\times10^{-2}$ & $5.65552\times10^{-1}$ & $5.01669\times10^{-3}$ & 1.62356 \\
$2.0$ & $1.62585\times10^{-1}$  & $1.28444\times10^{-1}$ & $1.63300\times10^{0}$ & $7.86552\times10^{-2}$ & $6.14252\times10^{-1}$ & $4.81854\times10^{-1}$ & $4.16754\times10^{-1}$ & $4.44142\times10^{-2}$ & $5.65434\times10^{-1}$ & $8.39347\times10^{-4}$ & 1.62354 \\
$1.0$ & $3.27070\times10^{-1}$  & $8.86277\times10^{-2}$ & $1.14300\times10^{0}$ & $7.75395\times10^{-2}$ & $6.16612\times10^{-1}$ & $4.86021\times10^{-1}$ & $8.91191\times10^{-1}$ & $9.48500\times10^{-2}$ & $5.65059\times10^{-1}$ & $5.65015\times10^{-3}$ & 1.62347 \\
\hline
$0.5$ & $6.62445\times10^{-1}$  & $5.95204\times10^{-2}$ & $7.93001\times10^{-1}$ & $7.50572\times10^{-2}$ & $6.21927\times10^{-1}$ & $4.95258\times10^{-1}$ & $2.05635\times10^{0}$ & $2.18273\times10^{-1}$ & $5.64303\times10^{-1}$ & $3.66372\times10^{-3}$ & 1.62333 \\
$0.3$ & $1.12526\times10^{0}$  & $4.27849\times10^{-2}$ & $5.99001\times10^{-1}$ & $7.14271\times10^{-2}$ & $6.30406\times10^{-1}$ & $5.09562\times10^{-1}$ & $4.23693\times10^{0}$ & $4.48081\times10^{-1}$ & $5.63266\times10^{-1}$ &  $9.11920\times10^{-3}$ & 1.62311 \\
$0.2$ & $1.73635\times10^{0}$  & $3.14517\times10^{-2}$ & $4.77001\times10^{-1}$ & $6.59364\times10^{-2}$ & $6.44181\times10^{-1}$ & $5.31996\times10^{-1}$ & $8.65548\times10^{0}$ & $9.11403\times10^{-1}$ & $5.62044\times10^{-1}$ &  $7.05363\times10^{-3}$ & 1.62278 \\
\hline
$0.15$ & $2.39865\times10^{0}$  & $2.41230\times10^{-2}$ & $4.07001\times10^{-1}$ & $5.92702\times10^{-2}$ & $6.63548\times10^{-1}$ & $5.61946\times10^{-1}$ &  $1.68117\times10^{1}$ & $1.76500\times10^{0}$ & $5.61211\times10^{-1}$  &  $9.52828\times10^{-3}$ & 1.62239 \\
$0.14$ & $2.60081\times10^{0}$  & $2.24259\times10^{-2}$ & $3.93001\times10^{-1}$ & $5.70632\times10^{-2}$ & $6.70637\times10^{-1}$ & $5.72538\times10^{-1}$ & $2.03949\times10^{1}$ & $2.14055\times10^{0}$ & $5.61128\times10^{-1}$  &  $1.01614\times10^{-2}$ & 1.62225 \\
$0.13$ & $2.84350\times10^{0}$  & $2.06203\times10^{-2}$ & $3.80001\times10^{-1}$ & $5.42638\times10^{-2}$ & $6.80073\times10^{-1}$ & $5.86379\times10^{-1}$ & $2.56589\times10^{1}$ & $2.69362\times10^{0}$ & $5.61190\times10^{-1}$  &  $8.83738\times10^{-3}$ & 1.62209 \\
$0.12$ & $3.14240\times10^{0}$  & $1.86835\times10^{-2}$ & $3.69001\times10^{-1}$ & $5.06326\times10^{-2}$ & $6.93337\times10^{-1}$ & $6.05343\times10^{-1}$ & $3.40099\times10^{1}$ & $3.57521\times10^{0}$ & $5.61577\times10^{-1}$  &  $5.28167\times10^{-3}$ & 1.62188 \\
$0.11$ & $3.52490\times10^{0}$  & $1.65825\times10^{-2}$ & $3.62001\times10^{-1}$ & $4.58080\times10^{-2}$ & $7.13594\times10^{-1}$ & $6.33260\times10^{-1}$ & $4.88977\times10^{1}$ & $5.16148\times10^{0}$ & $5.62734\times10^{-1}$  &  $6.38807\times10^{-3}$ & 1.62159 \\
\hline
$0.1$ & $4.05150\times10^{0}$  & $1.42556\times10^{-2}$ & $3.68001\times10^{-1}$ & $3.87378\times10^{-2}$ & $7.50107\times10^{-1}$ & $6.81035\times10^{-1}$ & $8.17923\times10^{1}$ & $8.74006\times10^{0}$ & $5.66189\times10^{-1}$  &  $8.76561\times10^{-3}$ & 1.62117 \\
\hline
$0.09$ & $5.08620\times10^{0}$  & $1.13990\times10^{-2}$ & $4.83001\times10^{-1}$ & $2.36004\times10^{-2}$ & $8.89250\times10^{-1}$ & $8.44163\times10^{-1}$ & $2.57654\times10^{2}$ & $2.96813\times10^{1}$ & $5.87872\times10^{-1}$  &  $5.72277\times10^{-3}$ & 1.62023 \\
[0.5ex]
  \hline\hline
 \end{tabular}
 }

 \label{table:MB}
\end{table}

\begin{table}
  \caption{Marginally stable configurations at different boundary temperatures $b = k_B T(R) / m c^2$ for a Fermi--Dirac distribution with $\alpha(R) =-10$. No marginally stable configurations exist for $b=0.08$ and $0.05$. 
  }
 \centering
 %\begin{ruledtabular}
 \resizebox{\columnwidth}{!}{
 \begin{tabular}{l l l l l l l l l l l l}
  \hline\hline
  % inserts table 
  %heading
  $b$  &  $w(0)$  & $\hat{M}$ & $\hat{R}$ & $C=\hat{M}/\hat{R}$ & $Z(0)$ & $\epsilon_c(0)/mc^2$ & $\hat{\rho}(0)$ & $\hat{p}(0)$ & $v(0)/c$ & $v(R)$ & $\langle\gamma\rangle=\gamma_{\text{cr}}$ \\ [0.5ex] 
  \hline\hline
$5.0$ & $6.48150\times10^{-2}$  & $2.06020\times10^{-1}$ & $2.59700\times10^{0}$ & $7.93298\times10^{-2}$ & $6.12906\times10^{-1}$ & $4.79447\times10^{-1}$ & $1.60337\times10^{-1}$ & $1.71005\times10^{-2}$ & $5.65651\times10^{-1}$ & $3.62915\times10^{-3}$ & 1.62358 \\
$3.0$ & $1.08185\times10^{-1}$  & $1.58577\times10^{-1}$ & $2.00600\times10^{0}$ & $7.90511\times10^{-2}$ & $6.13491\times10^{-1}$ & $4.80490\times10^{-1}$ & $2.71846\times10^{-1}$ & $2.89833\times10^{-2}$ & $5.65552\times10^{-1}$ & $5.19761\times10^{-3}$ & 1.62356 \\
$2.0$ & $1.62585\times10^{-1}$  & $1.28447\times10^{-1}$ & $1.63300\times10^{0}$ & $7.86571\times10^{-2}$ & $6.14253\times10^{-1}$ & $4.81853\times10^{-1}$ & $4.16733\times10^{-1}$ & $4.44120\times10^{-2}$ & $5.65434\times10^{-1}$ & $1.59714\times10^{-3}$ & 1.62354 \\
$1.0$ & $3.27070\times10^{-1}$  & $8.86301\times10^{-2}$ & $1.14300\times10^{0}$ & $7.75416\times10^{-2}$ & $6.16613\times10^{-1}$ & $4.86020\times10^{-1}$ & 
$8.91143\times10^{-1}$ & $9.48450\times10^{-2}$ & $5.65059\times10^{-1}$ & $5.81049\times10^{-3}$ & 1.62347 \\
\hline
$0.5$ & $6.62440\times10^{-1}$  & $5.95226\times10^{-2}$ & $7.93001\times10^{-1}$ & $7.50600\times10^{-2}$ & $6.21922\times10^{-1}$ & $4.95251\times10^{-1}$ & $2.05615\times10^{0}$ & $2.18251\times10^{-1}$ & $5.64302\times10^{-1}$ & $3.96802\times10^{-3}$ & 1.62333 \\
$0.3$ & $1.12526\times10^{0}$  & $4.27867\times10^{-2}$ & $5.99001\times10^{-1}$ & $7.14300\times10^{-2}$ & $6.30409\times10^{-1}$ & $5.09561\times10^{-1}$ & $4.23656\times10^{0}$ & $4.48046\times10^{-1}$ & $5.63268\times10^{-1}$ & $9.20451\times10^{-3}$ & 1.62311 \\
$0.2$ & $1.73630\times10^{0}$  & $3.14543\times10^{-2}$ & $4.77001\times10^{-1}$ & $6.59418\times10^{-2}$ & $6.44161\times10^{-1}$ & $5.31971\times10^{-1}$ & $8.65321\times10^{0}$ & $9.11156\times10^{-1}$ & $5.62042\times10^{-1}$ &  $7.22504\times10^{-3}$ & 1.62278 \\
\hline
$0.15$ & $2.39855\times10^{0}$  & $2.41259\times10^{-2}$ & $4.07001\times10^{-1}$ & $5.92774\times10^{-2}$ & $6.63521\times10^{-1}$ & $5.61909\times10^{-1}$ & 
$1.68047\times10^{1}$ & $1.76427\times10^{0}$ & $5.61213\times10^{-1}$  &  $9.57081\times10^{-3}$ & 1.62239 \\
$0.14$ & $2.60070\times10^{0}$  & $2.24288\times10^{-2}$ & $3.93001\times10^{-1}$ & $5.70707\times10^{-2}$ & $6.70611\times10^{-1}$ & $5.72501\times10^{-1}$ & 
$2.03854\times10^{1}$ & $2.13959\times10^{0}$ & $5.61133\times10^{-1}$  &  $1.01503\times10^{-2}$ & 1.62226 \\
$0.13$ & $2.84330\times10^{0}$  & $2.06239\times10^{-2}$ & $3.80001\times10^{-1}$ & $5.42732\times10^{-2}$ & $6.80023\times10^{-1}$ & $5.86315\times10^{-1}$ & $2.56416\times10^{1}$ & $2.69183\times10^{0}$ & $5.61193\times10^{-1}$  &  $8.77667\times10^{-3}$ & 1.62209 \\
$0.12$ & $3.14215\times10^{0}$  & $1.86872\times10^{-2}$ & $3.69001\times10^{-1}$ & $5.06427\times10^{-2}$ & $6.93282\times10^{-1}$ & $6.05269\times10^{-1}$ & $3.39819\times10^{1}$ & $3.57239\times10^{0}$ & $5.61586\times10^{-1}$  &  $4.92349\times10^{-3}$ & 1.62188 \\
$0.11$ & $3.52430\times10^{0}$  & $1.65879\times10^{-2}$ & $3.62001\times10^{-1}$ & $4.58227\times10^{-2}$ & $7.13454\times10^{-1}$ & $6.33090\times10^{-1}$ & 
$4.88239\times10^{1}$ & $5.15369\times10^{0}$ & $5.62735\times10^{-1}$  &  $5.71357\times10^{-3}$ & 1.62159 \\
\hline
$0.1$ & $4.05000\times10^{0}$  & $1.42556\times10^{-2}$ & $3.68001\times10^{-1}$ & $3.87591\times10^{-2}$ & $7.49753\times10^{-1}$ & $6.80629\times10^{-1}$ & $8.15368\times10^{1}$ & $8.71221\times10^{0}$ & $5.66172\times10^{-1}$  &  $7.35478\times10^{-3}$ & 1.62117 \\
\hline
$0.09$ & $5.06130\times10^{0}$  & $1.14308\times10^{-2}$ & $4.76001\times10^{-1}$ & $2.40141\times10^{-2}$ & $8.82317\times10^{-1}$ & $8.36583\times10^{-1}$ & 
$2.48175\times10^{2}$ & $2.84782\times10^{1}$ & $5.86729\times10^{-1}$  &  $4.63031\times10^{-3}$ & 1.62027 \\
$0.08$ & {------------------} &  {------------------} &  {------------------} &  {------------------} &  {------------------} &  {------------------} & 
 {------------------} &  {------------------} &  {------------------} &  {------------------} & {---------} \\
$0.05$ &  {------------------}  &  {------------------} &  {------------------} &  {------------------} &  {------------------} &  {------------------} & 
 {------------------} &  {------------------} &  {------------------} &  {------------------} & {---------} \\
\hline
$0.014$ & $2.60950\times10^{1}$  & $2.47115\times10^{-4}$ & $5.18000\times10^{-2}$ & $4.77055\times10^{-3}$ & $5.83182\times10^{-1}$ & $5.75618\times10^{-1}$ & 
$8.73487\times10^{4}$ & $9.65477\times10^{3}$ & $5.75842\times10^{-1}$  &  $1.94373\times10^{-3}$ & 1.62470 \\
$0.013$ & $2.75480\times10^{1}$  & $2.43927\times10^{-4}$ & $1.77000\times10^{-2}$ & $1.37812\times10^{-2}$ & $5.79839\times10^{-1}$ & $5.57927\times10^{-1}$ & 
$8.73394\times10^{4}$ & $9.63790\times10^{3}$ & $5.75369\times10^{-1}$  &  $2.85818\times10^{-3}$ & 1.62480 \\
$0.012$ & $2.92505\times10^{1}$  & $2.41484\times10^{-4}$ & $1.05000\times10^{-2}$ & $2.29984\times10^{-2}$ & $5.77219\times10^{-1}$ & $5.40731\times10^{-1}$ & 
$8.74315\times10^{4}$ & $9.63809\times10^{3}$ & $5.75072\times10^{-1}$  &  $1.34216\times10^{-2}$ & 1.62489 \\
$0.011$ & $3.12650\times10^{1}$  & $2.39448\times10^{-4}$ & $7.50001\times10^{-3}$ & $3.19264\times10^{-2}$ & $5.75073\times10^{-1}$ & $5.24112\times10^{-1}$ & 
$8.75621\times10^{4}$ & $9.64569\times10^{3}$ & $5.74869\times10^{-1}$  &  $1.14238\times10^{-2}$ & 1.62497 \\
\hline
$10^{-2}$ & $3.36825\times10^{1}$  & $2.37725\times10^{-4}$ & $5.80001\times10^{-3}$ & $4.09870\times10^{-2}$ & $5.73283\times10^{-1}$ & $5.07737\times10^{-1}$ & 
$8.77049\times10^{4}$ & $9.65655\times10^{3}$ & $5.74725\times10^{-1}$  &  $1.63905\times10^{-2}$ & 1.62504 \\
$10^{-3}$ & $2.72040\times10^{2}$  & $2.31026\times10^{-4}$ & $1.90001\times10^{-3}$ & $1.21592\times10^{-1}$ & $5.66625\times10^{-1}$ & $3.70760\times10^{-1}$ & $8.85156\times10^{4}$ & $9.73440\times10^{3}$ & $5.74388\times10^{-1}$  &  $5.49270\times10^{-2}$ & 1.62534 \\
$10^{-4}$ & $2.65480\times10^{3}$  & $2.30790\times10^{-4}$ & $1.80001\times10^{-3}$ & $1.28216\times10^{-1}$ & $5.66484\times10^{-1}$ & $3.58606\times10^{-1}$ & $8.85619\times10^{4}$ & $9.74112\times10^{3}$ & $5.74436\times10^{-1}$  &  $9.06083\times10^{-2}$ & 1.62533 \\
$10^{-5}$ & $2.64762\times10^{4}$  & $2.30707\times10^{-4}$ & $1.80001\times10^{-3}$ & $1.28170\times10^{-1}$ & $5.66413\times10^{-1}$ & $3.57643\times10^{-1}$ & $8.85773\times10^{4}$ & $9.74346\times10^{3}$ & $5.74455\times10^{-1}$  &  $8.99450\times10^{-2}$ & 1.62532 \\
[0.5ex]
  \hline\hline
 \end{tabular}
 }
 
 \label{table:FDm10}
\end{table}

\begin{table}
  \caption{Marginally stable configurations at different boundary temperatures $b = k_B T(R) / m c^2$ for a Fermi--Dirac distribution with $\alpha(R) = -5$.
  }
 \centering
 %\begin{ruledtabular}
 \resizebox{\columnwidth}{!}{
 \begin{tabular}{l l l l l l l l l l l l}
  \hline\hline
  % inserts table 
  %heading
  $b$  &  $w(0)$  & $\hat{M}$ & $\hat{R}$ & $C=\hat{M}/\hat{R}$ & $Z(0)$ & $\epsilon_c(0)/mc^2$ & $\hat{\rho}(0)$ & $\hat{p}(0)$ & $v(0)/c$ & $v(R)/c$ & $\langle\gamma\rangle=\gamma_{\text{cr}}$ \\ [0.5ex] 
  \hline\hline
$5.0$ & $6.48140\times10^{-2}$  & $2.06732\times10^{-1}$ & $2.60600\times10^{0}$ & $7.93293\times10^{-2}$ & $6.12906\times10^{-1}$ & $4.79440\times10^{-1}$ & $1.59225\times10^{-1}$ & $1.69823\times10^{-2}$ & $5.65657\times10^{-1}$ & $2.49564\times10^{-3}$ & 1.62358 \\
$3.0$ & $1.08180\times10^{-1}$  & $1.59140\times10^{-1}$ & $2.01300\times10^{0}$ & $7.90562\times10^{-2}$ & $6.13475\times10^{-1}$ & $4.80460\times10^{-1}$ & $2.69885\times10^{-1}$ & $2.87747\times10^{-2}$ & $5.65557\times10^{-1}$ & $4.66725\times10^{-3}$ & 1.62356 \\
$2.0$ & $1.62570\times10^{-1}$  & $1.28920\times10^{-1}$ & $1.63800\times10^{0}$ & $7.87056\times10^{-2}$ & $6.14210\times10^{-1}$ & $4.81768\times10^{-1}$ & $4.13567\times10^{-1}$ & $4.40748\times10^{-2}$ & $5.65435\times10^{-1}$ & $6.03557\times10^{-3}$ & 1.62354 \\
$1.0$ & $3.27010\times10^{-1}$  & $8.89895\times10^{-2}$ & $1.14700\times10^{0}$ & $7.75845\times10^{-2}$ & $6.16534\times10^{-1}$ & $4.85882\times10^{-1}$ & 
$8.83448\times10^{-1}$ & $9.40293\times10^{-2}$ & $5.65069\times10^{-1}$ & $6.55257\times10^{-3}$ & 1.62347 \\
\hline
$0.5$ & $6.62150\times10^{-1}$  & $5.98192\times10^{-2}$ & $7.96001\times10^{-1}$ & $7.51496\times10^{-2}$ & $6.21728\times10^{-1}$ & $4.94923\times10^{-1}$ & $2.03283\times10^{0}$ & $2.15800\times10^{-1}$ & $5.64334\times10^{-1}$ & $4.76202\times10^{-3}$ & 1.62333 \\
$0.3$ & $1.12412\times10^{0}$  & $4.30752\times10^{-2}$ & $6.02001\times10^{-1}$ & $7.15534\times10^{-2}$ & $6.29916\times10^{-1}$ & $5.08818\times10^{-1}$ & $4.16478\times10^{0}$ & $4.40566\times10^{-1}$ & $5.63339\times10^{-1}$ & $4.99364\times10^{-3}$ & 1.62312 \\
$0.2$ & $1.73222\times10^{0}$  & $3.17787\times10^{-2}$ & $4.79001\times10^{-1}$ & $6.63436\times10^{-2}$ & $6.42888\times10^{-1}$ & $5.30065\times10^{-1}$ & $8.39869\times10^{0}$ & $8.84912\times10^{-1}$ & $5.62218\times10^{-1}$ & $6.48972\times10^{-3}$ & 1.62281 \\
\hline
$0.15$ & $2.38570\times10^{0}$  & $2.45354\times10^{-2}$ & $4.08001\times10^{-1}$ & $6.01358\times10^{-2}$ & $6.60194\times10^{-1}$ & $5.57237\times10^{-1}$ & 
$1.58716\times10^{1}$ & $1.66837\times10^{0}$ & $5.61560\times10^{-1}$  &  $8.15146\times10^{-3}$ & 1.62245 \\
$0.14$ & $2.58305\times10^{0}$  & $2.28743\times10^{-2}$ & $3.93001\times10^{-1}$ & $5.82041\times10^{-2}$ & $6.66221\times10^{-1}$ & $5.66395\times10^{-1}$ & $1.90079\times10^{1}$ & $1.99796\times10^{0}$ & $5.61548\times10^{-1}$  &  $1.15083\times10^{-2}$ & 1.62234 \\
$0.13$ & $2.81771\times10^{0}$  & $2.11203\times10^{-2}$ & $3.79001\times10^{-1}$ & $5.57262\times10^{-2}$ & $6.73895\times10^{-1}$ & $5.77974\times10^{-1}$ & $2.34430\times10^{1}$ & $2.46531\times10^{0}$ & $5.61681\times10^{-1}$  &  $9.85743\times10^{-3}$ & 1.62220 \\
$0.12$ & $3.10200\times10^{0}$  & $1.92611\times10^{-2}$ & $3.66001\times10^{-1}$ & $5.26259\times10^{-2}$ & $6.83976\times10^{-1}$ & $5.62103\times10^{-1}$ & $3.00599\times10^{1}$ & $3.16591\times10^{0}$ & $5.67382\times10^{-1}$  &  $6.30871\times10^{-3}$ & 1.62358 \\
$0.11$ & $3.45460\times10^{0}$  & $1.72843\times10^{-2}$ & $3.54001\times10^{-1}$ & $4.88255\times10^{-2}$ & $6.97738\times10^{-1}$ & $6.12842\times10^{-1}$ & 
$4.06396\times10^{1}$ & $4.29571\times10^{0}$ & $5.63123\times10^{-1}$  &  $1.03703\times10^{-2}$ & 1.62182 \\
\hline
$0.1$ & $3.90455\times10^{0}$  & $1.51841\times10^{-2}$ & $3.47001\times10^{-1}$ & $4.37581\times10^{-2}$ & $7.17285\times10^{-1}$ & $6.40510\times10^{-1}$ & $5.91033\times10^{1}$ & $6.29776\times10^{0}$ & $5.65390\times10^{-1}$  &  $8.80445\times10^{-3}$ & 1.62155 \\
\hline
$0.09$ & $4.49800\times10^{0}$  & $1.29781\times10^{-2}$ & $3.50001\times10^{-1}$ & $3.70802\times10^{-2}$ & $7.46112\times10^{-1}$ & $6.80146\times10^{-1}$ & 
$9.49222\times10^{1}$ & $1.02939\times10^{1}$ & $5.70384\times10^{-1}$  &  $4.80109\times10^{-3}$ & 1.62124 \\
$0.08$ & $5.29500\times10^{0}$  & $1.07669\times10^{-2}$ & $3.74001\times10^{-1}$ & $2.87884\times10^{-2}$ & $7.87032\times10^{-1}$ & $7.34871\times10^{-1}$ & 
$1.69995\times10^{2}$ & $1.91071\times10^{1}$ & $5.80683\times10^{-1}$  &  $6.39985\times10^{-3}$ & 1.62092 \\
$0.05$ & $8.47930\times10^{0}$  & $5.68169\times10^{-3}$ & $3.95001\times10^{-1}$ & $1.43840\times10^{-2}$ & $7.61525\times10^{-1}$ & $7.36004\times10^{-1}$ & 
$5.22206\times10^{2}$ & $6.23475\times10^{1}$ & $5.98479\times10^{-1}$  &  $8.27190\times10^{-4}$ & 1.62164 \\
\hline
$10^{-2}$ & $3.00295\times10^{1}$  & $2.87811\times10^{-3}$ & $3.3001\times10^{-2}$ & $8.72129\times10^{-2}$ & $5.71331\times10^{-1}$ & $4.28740\times10^{-1}$ & 
$5.90467\times10^{2}$ & $6.49386\times10^{1}$ & $5.74400\times10^{-1}$  &  $2.93362\times10^{-2}$ & 1.62511 \\
$10^{-3}$ & $2.68270\times10^{2}$  & $2.81430\times10^{-3}$ & $2.3001\times10^{-2}$ & $1.22355\times10^{-1}$ & $5.66349\times10^{-1}$ & $3.65189\times10^{-1}$ & $5.95686\times10^{2}$ & $6.54783\times10^{1}$ & $5.74250\times10^{-1}$  &  $4.96081\times10^{-2}$ & 1.62534 \\
$10^{-4}$ & $2.65000\times10^{3}$  & $2.81169\times10^{-3}$ & $2.2001\times10^{-2}$ & $1.27798\times10^{-1}$ & $5.66258\times10^{-1}$ & $3.58108\times10^{-1}$ & $5.96047\times10^{2}$ & $6.55313\times10^{1}$ & $5.74308\times10^{-1}$  &  $8.67470\times10^{-2}$ & 1.62533 \\
$10^{-5}$ & $2.64620\times10^{4}$  & $2.81155\times10^{-3}$ & $2.2001\times10^{-2}$ & $1.27792\times10^{-1}$ & $5.66215\times10^{-1}$ & $3.57587\times10^{-1}$ & $5.95992\times10^{2}$ & $6.55228\times10^{1}$ & $5.74297\times10^{-1}$  &  $8.63942\times10^{-2}$ & 1.62533 \\
[0.5ex]
  \hline\hline
 \end{tabular}
 }
 
 \label{table:FDm5}
\end{table}

\begin{table}
  \caption{Marginally stable configurations at different boundary temperatures $b = k_B T(R) / m c^2$ for a Fermi--Dirac distribution with $\alpha(R) = -1$.}
 \centering
 %\begin{ruledtabular}
 \resizebox{\columnwidth}{!}{
 \begin{tabular}{l l l l l l l l l l l l}
  \hline\hline
  % inserts table 
  %heading
  $b$  &  $w(0)$  & $\hat{M}$ & $\hat{R}$ & $C=\hat{M}/\hat{R}$ & $Z(0)$ & $\epsilon_c(0)/mc^2$ & $\hat{\rho}(0)$ & $\hat{p}(0)$ & $v(0)/c$ & $v(R)/c$ & $\langle\gamma\rangle=\gamma_{\text{cr}}$ \\ [0.5ex] 
  \hline\hline
$5.0$ & $6.47360\times10^{-2}$  & $2.42184\times10^{-1}$ & $3.04400\times10^{0}$ & $7.95612\times10^{-2}$ & $6.12415\times10^{-1}$ & $4.78585\times10^{-1}$ & $1.15599\times10^{-1}$ & $1.23323\times10^{-2}$ & $5.65724\times10^{-1}$ & $3.00356\times10^{-3}$ & 1.62359 \\
$3.0$ & $1.07963\times10^{-1}$  & $1.87069\times10^{-1}$ & $2.35500\times10^{0}$ & $7.94347\times10^{-2}$ & $6.12668\times10^{-1}$ & $4.79041\times10^{-1}$ & $1.94126\times10^{-1}$ & $2.07066\times10^{-2}$ & $5.65684\times10^{-1}$ & $3.24415\times10^{-3}$ & 1.62358 \\
$2.0$ & $1.62070\times10^{-1}$  & $1.52214\times10^{-1}$ & $1.92000\times10^{0}$ & $7.92780\times10^{-2}$ & $6.12965\times10^{-1}$ & $4.79590\times10^{-1}$ & $2.93887\times10^{-1}$ & $3.13420\times10^{-2}$ & $5.65632\times10^{-1}$ & $3.37873\times10^{-3}$ & 1.62357 \\
$1.0$ & $3.24880\times10^{-1}$  & $1.06550\times10^{-1}$ & $1.35100\times10^{0}$ & $7.88676\times10^{-2}$ & $6.13870\times10^{-1}$ & $4.81189\times10^{-1}$ & 
$6.03788\times10^{-1}$ & $6.43667\times10^{-2}$ & $5.65521\times10^{-1}$ & $7.20175\times10^{-3}$ & 1.62355 \\
\hline
$0.5$ & $6.52250\times10^{-1}$  & $7.39634\times10^{-2}$ & $9.48001\times10^{-1}$ & $7.80204\times10^{-2}$ & $6.15322\times10^{-1}$ & $4.83954\times10^{-1}$ & $1.26653\times10^{0}$ & $1.34949\times10^{-1}$ & $5.65375\times10^{-1}$ & $1.18484\times10^{-3}$ & 1.62351 \\
$0.3$ & $1.09082\times10^{0}$  & $5.61277\times10^{-2}$ & $7.26001\times10^{-1}$ & $7.73107\times10^{-2}$ & $6.16473\times10^{-1}$ & $4.86372\times10^{-1}$ & $2.21714\times10^{0}$ & $2.36252\times10^{-1}$ & $5.65395\times10^{-1}$ & $9.89832\times10^{-3}$ & 1.62347 \\
$0.2$ & $1.63855\times10^{0}$  & $4.50009\times10^{-2}$ & $5.86001\times10^{-1}$ & $7.67932\times10^{-2}$ & $6.16588\times10^{-1}$ & $4.87394\times10^{-1}$ & $3.44629\times10^{0}$ & $3.67762\times10^{-1}$ & $5.65807\times10^{-1}$ & $1.01965\times10^{-2}$ & 1.62347 \\
\hline
$0.15$ & $2.18075\times10^{0}$  & $3.86021\times10^{-2}$ & $5.03001\times10^{-1}$ & $7.67436\times10^{-2}$ & $6.15194\times10^{-1}$ & $4.86114\times10^{-1}$ & 
$4.62583\times10^{0}$ & $4.94868\times10^{-1}$ & $5.66514\times10^{-1}$ & $5.65521\times10^{-3}$ & 1.62351 \\
$0.14$ & $2.33410\times10^{0}$  & $3.72481\times10^{-2}$ & $4.84001\times10^{-1}$ & $7.69588\times10^{-2}$ & $6.14570\times10^{-1}$ & $4.85301\times10^{-1}$ & $4.94221\times10^{0}$ & $5.29158\times10^{-1}$ & $5.66751\times10^{-1}$ & $1.22325\times10^{-2}$ & 1.62353 \\
$0.13$ & $2.51005\times10^{0}$  & $3.58709\times10^{-2}$ & $4.66001\times10^{-1}$ & $7.69760\times10^{-2}$ & $6.13756\times10^{-1}$ & $4.84352\times10^{-1}$ & $5.29293\times10^{0}$ & $5.67283\times10^{-1}$ & $5.67038\times10^{-1}$ & $2.12217\times10^{-3}$ & 1.62355 \\
$0.12$ & $2.71390\times10^{0}$  & $3.44727\times10^{-2}$ & $4.46001\times10^{-1}$ & $7.72929\times10^{-2}$ & $6.12694\times10^{-1}$ & $4.82905\times10^{-1}$ & $5.68106\times10^{0}$ & $6.09621\times10^{-1}$ & $5.67382\times10^{-1}$ & $8.82191\times10^{-3}$ & 1.62358 \\
$0.11$ & $2.95270\times10^{0}$  & $3.30576\times10^{-2}$ & $4.26001\times10^{-1}$ & $7.75998\times10^{-2}$ & $6.11298\times10^{-1}$ & $4.81022\times10^{-1}$ & 
$6.10855\times10^{0}$ & $6.56424\times10^{-1}$ & $5.67785\times10^{-1}$ & $5.00084\times10^{-3}$ & 1.62362 \\
\hline
$0.1$ & $3.23630\times10^{0}$  & $3.16303\times10^{-2}$ & $4.05001\times10^{-1}$ & $7.80994\times10^{-2}$ & $6.09503\times10^{-1}$ & $4.78477\times10^{-1}$ & $6.57701\times10^{0}$ & $7.07950\times10^{-1}$ & $5.68261\times10^{-1}$ & $2.73960\times10^{-3}$ & 1.62367 \\
\hline
$0.09$ & $3.57870\times10^{0}$  & $3.01981\times10^{-2}$ & $3.83001\times10^{-1}$ & $7.88460\times10^{-2}$ & $6.07229\times10^{-1}$ & $4.75096\times10^{-1}$ & $7.08595\times10^{0}$ & $7.64245\times10^{-1}$ & $5.68824\times10^{-1}$ & $4.49398\times10^{-3}$ & 1.62374 \\
$0.08$ & $4.00040\times10^{0}$  & $2.87721\times10^{-2}$ & $3.60001\times10^{-1}$ & $7.99223\times10^{-2}$ & $6.04347\times10^{-1}$ & $4.70621\times10^{-1}$ & $7.63014\times10^{0}$ & $8.24800\times10^{-1}$ & $5.69467\times10^{-1}$ &  $8.11910\times10^{-3}$ & 1.62383 \\
$0.05$ & $6.18610\times10^{0}$  & $2.47018\times10^{-2}$ & $2.87001\times10^{-1}$ & $8.60687\times10^{-2}$ & $5.91211\times10^{-1}$ & $4.47810\times10^{-1}$ & 
$9.34675\times10^{0}$ & $1.01863\times10^{0}$ & $5.71791\times10^{-1}$  &  $3.66863\times10^{-3}$ & 1.62427 \\
\hline
$10^{-2}$ & $2.76760\times10^{1}$  & $2.10095\times10^{-2}$ & $1.88001\times10^{-1}$ & $1.11752\times10^{-1}$ & $5.67938\times10^{-1}$ & $3.82386\times10^{-1}$ & 
$1.08281\times10^{1}$ & $1.18911\times10^{0}$ & $5.73978\times10^{-1}$  &  $2.50725\times10^{-2}$ & 1.62524 \\
$10^{-3}$ & $2.65650\times10^{2}$  & $2.07922\times10^{-2}$ & $1.70001\times10^{-1}$ & $1.22306\times10^{-1}$ & $5.65813\times10^{-1}$ & $3.61522\times10^{-1}$ & $1.08849\times10^{1}$ & $1.19536\times10^{0}$ & $5.73982\times10^{-1}$  &  $2.29994\times10^{-2}$ & 1.62534 \\
$10^{-4}$ & $2.64470\times10^{3}$  & $2.07897\times10^{-2}$ & $1.68001\times10^{-1}$ & $1.23748\times10^{-1}$ & $5.65793\times10^{-1}$ & $3.59224\times10^{-1}$ & $1.08856\times10^{1}$ & $1.19545\times10^{0}$ & $5.73984\times10^{-1}$  &  $3.09253\times10^{-2}$ & 1.62534 \\
$10^{-5}$ & $2.64340\times10^{4}$  & $2.07890\times10^{-2}$ & $1.68001\times10^{-1}$ & $1.23743\times10^{-1}$ & $5.65788\times10^{-1}$ & $3.59048\times10^{-1}$ & $1.08859\times10^{1}$ & $1.19549\times10^{0}$ & $5.73986\times10^{-1}$  &  $2.99647\times10^{-2}$ & 1.62534 \\
[0.5ex]
  \hline\hline
 \end{tabular}
 }
 
 \label{table:FDm1}
\end{table}

\begin{table}
  \caption{Marginally stable configurations at different boundary temperatures $b = k_B T(R) / m c^2$ for a Fermi–Dirac distribution with $\alpha(R) = 0$, corresponding to a completely degenerate core.
  }
 \centering
 %\begin{ruledtabular}
 \resizebox{\columnwidth}{!}{
 \begin{tabular}{l l l l l l l l l l l l}
  \hline\hline
  % inserts table 
  %heading
  $b$  &  $w(0)$  & $\hat{M}$ & $\hat{R}$ & $C=\hat{M}/\hat{R}$ & $Z(0)$ & $\epsilon_c(0)/mc^2$ & $\hat{\rho}(0)$ & $\hat{p}(0)$ & $v(0)/c$ & $v(R)/c$ & $\langle\gamma\rangle=\gamma_{\text{cr}}$ \\ [0.5ex] 
  \hline\hline
$5.0$ & $6.46710\times10^{-2}$  & $2.94115\times10^{-1}$ & $3.68700\times10^{0}$ & $7.97708\times10^{-2}$ & $6.12023\times10^{-1}$ & $4.77869\times10^{-1}$ & $7.81440\times10^{-2}$ & $8.33855\times10^{-3}$ & $5.65794\times10^{-1}$ & $4.40456\times10^{-3}$ & 1.62360 \\
$3.0$ & $1.07780\times10^{-1}$  & $2.27846\times10^{-1}$ & $2.85600\times10^{0}$ & $7.97778\times10^{-2}$ & $6.11999\times10^{-1}$ & $4.77836\times10^{-1}$ & $1.30191\times10^{-1}$ & $1.38924\times10^{-2}$ & $5.65796\times10^{-1}$ & $4.49651\times10^{-3}$ & 1.62360 \\
$2.0$ & $1.61655\times10^{-1}$  & $1.86075\times10^{-1}$ & $2.33200\times10^{0}$ & $7.97921\times10^{-2}$ & $6.11951\times10^{-1}$ & $4.77769\times10^{-1}$ & $1.95139\times10^{-1}$ & $2.08233\times10^{-2}$ & $5.65800\times10^{-1}$ & $4.73966\times10^{-3}$ & 1.62360 \\
$1.0$ & $3.23190\times10^{-1}$  & $1.31718\times10^{-1}$ & $1.64900\times10^{0}$ & $7.98772\times10^{-2}$ & $6.11796\times10^{-1}$ & $4.77498\times10^{-1}$ & $3.88896\times10^{-1}$ & $4.15081\times10^{-2}$ & $5.65861\times10^{-1}$ & $6.26828\times10^{-3}$ & 1.62361 \\
\hline
$0.5$ & $6.45350\times10^{-1}$  & $9.35447\times10^{-2}$ & $1.16700\times10^{0}$ & $8.01582\times10^{-2}$ & $6.11080\times10^{-1}$ & $4.76365\times10^{-1}$ & $7.66630\times10^{-1}$ & $8.18806\times10^{-2}$ & $5.66054\times10^{-1}$ & $7.43761\times10^{-3}$ & 1.62363 \\
$0.3$ & $1.07176\times10^{0}$  & $7.31815\times10^{-2}$ & $9.06001\times10^{-1}$ & $8.07742\times10^{-2}$ & $6.09507\times10^{-1}$ & $4.73861\times10^{-1}$ & $1.23686\times10^{0}$ & $1.32302\times10^{-1}$ & $5.66479\times10^{-1}$ & $8.46604\times10^{-3}$ & 1.62368 \\
$0.2$ & $1.59765\times10^{0}$  & $6.08360\times10^{-2}$ & $7.43001\times10^{-1}$ & $8.18788\times10^{-2}$ & $6.06833\times10^{-1}$ & $4.69514\times10^{-1}$ & $1.75163\times10^{0}$ & $1.87835\times10^{-1}$ & $5.67189\times10^{-1}$ & $1.03413\times10^{-2}$ & 1.62376 \\
\hline
$0.15$ & $2.11415\times10^{0}$  & $5.38941\times10^{-2}$ & $6.48001\times10^{-1}$ & $8.31698\times10^{-2}$ & $6.03689\times10^{-1}$ & $4.64346\times10^{-1}$ & $2.17671\times10^{0}$ & $2.34054\times10^{-1}$ & $5.67961\times10^{-1}$ & $9.15085\times10^{-3}$ & 1.62385 \\
$0.14$ & $2.26000\times10^{0}$  & $5.24381\times10^{-2}$ & $6.28001\times10^{-1}$ & $8.35000\times10^{-2}$ & $6.02774\times10^{-1}$ & $4.62840\times10^{-1}$ & $2.28244\times10^{0}$ & $2.45619\times10^{-1}$ & $5.68188\times10^{-1}$ & $2.48344\times10^{-3}$ & 1.62388 \\
$0.13$ & $2.42735\times10^{0}$  & $5.09607\times10^{-2}$ & $6.06001\times10^{-1}$ & $8.40934\times10^{-2}$ & $6.01719\times10^{-1}$ & $4.60972\times10^{-1}$ & $2.39629\times10^{0}$ & $2.58106\times10^{-1}$ & $5.68448\times10^{-1}$ &  $1.11196\times10^{-2}$ & 1.62392 \\
$0.12$ & $2.62130\times10^{0}$  & $4.94640\times10^{-2}$ & $5.85001\times10^{-1}$ & $8.45536\times10^{-2}$ & $6.00480\times10^{-1}$ & $4.58903\times10^{-1}$ & $2.51850\times10^{0}$ & $2.71548\times10^{-1}$ & $5.68739\times10^{-1}$  &  $3.25580\times10^{-3}$ & 1.62396 \\
$0.11$ & $2.84880\times10^{0}$  & $4.79502\times10^{-2}$ & $5.62001\times10^{-1}$ & $8.53205\times10^{-2}$ & $5.99026\times10^{-1}$ & $4.56333\times10^{-1}$ & $2.64931\times10^{0}$ & $2.85982\times10^{-1}$ & $5.69068\times10^{-1}$ & $9.85064\times10^{-3}$ & 1.62400 \\
\hline
$0.1$ & $3.11960\times10^{0}$  & $4.64223\times10^{-2}$ & $5.39001\times10^{-1}$ & $8.61266\times10^{-2}$ & $5.97349\times10^{-1}$ & $4.53364\times10^{-1}$ & $2.78900\times10^{0}$ & $3.01466\times10^{-1}$ & $5.69450\times10^{-1}$ & $8.63863\times10^{-3}$ & 1.62406 \\
\hline
$0.09$ & $3.44745\times10^{0}$  & $4.48864\times10^{-2}$ & $5.15001\times10^{-1}$ & $8.71578\times10^{-2}$ & $5.95385\times10^{-1}$ & $4.49793\times10^{-1}$ & $2.93687\times10^{0}$ & $3.17934\times10^{-1}$ & $5.69884\times10^{-1}$ & $9.79655\times10^{-3}$ & 1.62413 \\
$0.08$ & $3.85285\times10^{0}$  & $4.33505\times10^{-2}$ & $4.90001\times10^{-1}$ & $8.84703\times10^{-2}$ & $5.93079\times10^{-1}$ & $4.45477\times10^{-1}$ & 
$3.09161\times10^{0}$ & $3.35258\times10^{-1}$ & $5.70371\times10^{-1}$ & $1.28836\times10^{-2}$ & 1.62421 \\
$0.05$ & $5.98040\times10^{0}$  & $3.88844\times10^{-2}$ & $4.12001\times10^{-1}$ & $9.43795\times10^{-2}$ & $5.83694\times10^{-1}$ & $4.26527\times10^{-1}$ & 
$3.56856\times10^{0}$ & $3.89383\times10^{-1}$ & $5.72141\times10^{-1}$ & $9.74573\times10^{-3}$ & 1.62455 \\
\hline
$10^{-2}$ & $2.73890\times10^{1}$  & $3.45579\times10^{-2}$ & $3.03001\times10^{-1}$ & $1.14052\times10^{-1}$ & $5.67411\times10^{-1}$ & $3.77171\times10^{-1}$ & 
$3.98720\times10^{0}$ & $4.37833\times10^{-1}$ & $5.73959\times10^{-1}$ & $8.36024\times10^{-3}$ & 1.62526 \\
$10^{-3}$ & $2.65345\times10^{2}$  & $3.42798\times10^{-2}$ & $2.80001\times10^{-1}$ & $1.22427\times10^{-1}$ & $5.65797\times10^{-1}$ & $3.61045\times10^{-1}$ & $4.00419\times10^{0}$ & $4.39725\times10^{-1}$ & $5.73976\times10^{-1}$ & $1.81042\times10^{-2}$ & 1.62534 \\
$10^{-4}$ & $2.64440\times10^{3}$  & $3.42765\times10^{-2}$ & $2.78001\times10^{-1}$ & $1.23296\times10^{-1}$ & $5.65796\times10^{-1}$ & $3.59399\times10^{-1}$ & $4.00461\times10^{0}$ & $4.39783\times10^{-1}$ & $5.73984\times10^{-1}$ & $1.67281\times10^{-2}$ & 1.62535 \\
$10^{-5}$ & $2.64330\times10^{4}$  & $3.42758\times10^{-2}$ & $2.78001\times10^{-1}$ & $1.23294\times10^{-1}$ & $5.65767\times10^{-1}$ & $3.59244\times10^{-1}$ & $4.00429\times10^{0}$ & $4.39734\times10^{-1}$ & $5.73975\times10^{-1}$ & $1.37005\times10^{-2}$ & 1.62534 \\
[0.5ex]
  \hline\hline
 \end{tabular}
 }

 \label{table:FD0}
\end{table}

\begin{table}
  \caption{Marginally stable configurations at different boundary temperatures $b = k_B T(R) / (m c^2)$ for a Fermi–Dirac distribution with $\alpha(R) = 0.5$, corresponding to an overly degenerate core.
  }
 \centering
 %\begin{ruledtabular}
 \resizebox{\columnwidth}{!}{
 \begin{tabular}{l l l l l l l l l l l l}
  \hline\hline
  % inserts table 
  %heading
  $b$  &  $w(0)$  & $\hat{M}$ & $\hat{R}$ & $C=\hat{M}/\hat{R}$ & $Z(0)$ & $\epsilon_c(0)/mc^2$ & $\hat{\rho}(0)$ & $\hat{p}(0)$ & $v(0)/c$ & $v(R)/c$ & $\langle\gamma\rangle=\gamma_{\text{cr}}$ \\ [0.5ex] 
  \hline\hline
$5.0$ & $6.46370\times10^{-2}$  & $3.39241\times10^{-1}$ & $4.24800\times10^{0}$ & $7.98589\times10^{-2}$ & $6.11818\times10^{-1}$ & $4.77507\times10^{-1}$ & $5.86445\times10^{-2}$ & $6.25860\times10^{-3}$ & $5.65830\times10^{-1}$ & $1.67907\times10^{-3}$ & 1.62361 \\
$3.0$ & $1.07685\times10^{-1}$  & $2.63201\times10^{-1}$ & $3.29200\times10^{0}$ & $7.99518\times10^{-2}$ & $6.11652\times10^{-1}$ & $4.77213\times10^{-1}$ & $9.73051\times10^{-2}$ & $1.03853\times10^{-2}$ & $5.65853\times10^{-1}$ & $4.59333\times10^{-3}$ & 1.62361 \\
$2.0$ & $1.61447\times10^{-1}$  & $2.15349\times10^{-1}$ & $2.69000\times10^{0}$ & $8.00554\times10^{-2}$ & $6.11458\times10^{-1}$ & $4.76859\times10^{-1}$ & $1.45130\times10^{-1}$ & $1.54919\times10^{-2}$ & $5.65893\times10^{-1}$ & $5.15083\times10^{-3}$ & 1.62362 \\
$1.0$ & $3.22360\times10^{-1}$  & $1.53287\times10^{-1}$ & $1.90700\times10^{0}$ & $8.03813\times10^{-2}$ & $6.10798\times10^{-1}$ & $4.75691\times10^{-1}$ & $2.84959\times10^{-1}$ & $3.04326\times10^{-2}$ & $5.66029\times10^{-1}$ & $5.86189\times10^{-3}$ & 1.62364 \\
\hline
$0.5$ & $6.42189\times10^{-1}$  & $1.10020\times10^{-1}$ & $1.35600\times10^{0}$ & $8.11353\times10^{-2}$ & $6.09199\times10^{-1}$ & $4.72927\times10^{-1}$ & $5.46316\times10^{-1}$ & $5.84123\times10^{-2}$ & $5.66357\times10^{-1}$ & $7.55503\times10^{-3}$ & 1.62368 \\
$0.3$ & $1.06375\times10^{0}$  & $8.71760\times10^{-2}$ & $1.06000\times10^{0}$ & $8.22414\times10^{-2}$ & $6.06723\times10^{-1}$ & $4.68683\times10^{-1}$ & $8.53315\times10^{-1}$ & $9.14147\times10^{-2}$ & $5.66910\times10^{-1}$ & $5.22698\times10^{-3}$ & 1.62376 \\
$0.2$ & $1.58197\times10^{0}$  & $7.34611\times10^{-2}$ & $8.76001\times10^{-1}$ & $8.38596\times10^{-2}$ & $6.03347\times10^{-1}$ & $4.62797\times10^{-1}$ & $1.16992\times10^{0}$ & $1.25675\times10^{-1}$ & $5.67685\times10^{-1}$ & $7.92247\times10^{-3}$ & 1.62386 \\
\hline
$0.15$ & $2.09042\times10^{0}$  & $6.57972\times10^{-2}$ & $7.69001\times10^{-1}$ & $8.55619\times10^{-2}$ & $5.99945\times10^{-1}$ & $4.56746\times10^{-1}$ & $1.41940\times10^{0}$ & $1.52895\times10^{-1}$ & $5.68467\times10^{-1}$ & $9.88243\times10^{-3}$ & 1.62397 \\
$0.14$ & $2.23398\times10^{0}$  & $6.41939\times10^{-2}$ & $7.46001\times10^{-1}$ & $8.60507\times10^{-2}$ & $5.98997\times10^{-1}$ & $4.55034\times10^{-1}$ & $1.47998\times10^{0}$ & $1.59541\times10^{-1}$ & $5.68682\times10^{-1}$ & $1.02912\times10^{-2}$ & 1.62400 \\
$0.13$ & $2.39875\times10^{0}$  & $6.25678\times10^{-2}$ & $7.23001\times10^{-1}$ & $8.65390\times10^{-2}$ & $5.97929\times10^{-1}$ & $4.53125\times10^{-1}$ & $1.54469\times10^{0}$ & $1.66661\times10^{-1}$ & $5.68927\times10^{-1}$ & $6.15021\times10^{-3}$ & 1.62404 \\
$0.12$ & $2.58980\times10^{0}$  & $6.09207\times10^{-2}$ & $6.99001\times10^{-1}$ & $8.71539\times10^{-2}$ & $5.96709\times10^{-1}$ & $4.50902\times10^{-1}$ & $1.61361\times10^{0}$ & $1.74264\times10^{-1}$ & $5.69199\times10^{-1}$ & $3.07272\times10^{-3}$ & 1.62408 \\
$0.11$ & $2.81410\times10^{0}$  & $5.92542\times10^{-2}$ & $6.74001\times10^{-1}$ & $8.79141\times10^{-2}$ & $5.95329\times10^{-1}$ & $4.48323\times10^{-1}$ & $1.68695\times10^{0}$ & $1.82383\times10^{-1}$ & $5.69512\times10^{-1}$ & $4.28778\times10^{-3}$ & 1.62413 \\
\hline
$0.1$ & $3.08114\times10^{0}$  & $5.75729\times10^{-2}$ & $6.48001\times10^{-1}$ & $8.88469\times10^{-2}$ & $5.93727\times10^{-1}$ & $4.45289\times10^{-1}$ & $1.76439\times10^{0}$ & $1.90986\times10^{-1}$ & $5.69855\times10^{-1}$ & $8.27089\times10^{-3}$ & 1.62418 \\
\hline
$0.09$ & $3.40490\times10^{0}$  & $5.58803\times10^{-2}$ & $6.22001\times10^{-1}$ & $8.98396\times10^{-2}$ & $5.91927\times10^{-1}$ & $4.41837\times10^{-1}$ & $1.84607\times10^{0}$ & $2.00109\times10^{-1}$ & $5.70256\times10^{-1}$ & $1.76092\times10^{-3}$ & 1.62425 \\
$0.08$ & $3.80560\times10^{0}$  & $5.41863\times10^{-2}$ & $5.94001\times10^{-1}$ & $9.12226\times10^{-2}$ & $5.89821\times10^{-1}$ & $4.37643\times10^{-1}$ & $1.93082\times10^{0}$ & $2.09615\times10^{-1}$ & $5.70692\times10^{-1}$ & $1.12090\times10^{-2}$ & 1.62432 \\
$0.05$ & $5.91600\times10^{0}$  & $4.92322\times10^{-2}$ & $5.08001\times10^{-1}$ & $9.69136\times10^{-2}$ & $5.81521\times10^{-1}$ & $4.20037\times10^{-1}$ & $2.19034\times10^{0}$ & $2.39117\times10^{-1}$ & $5.72282\times10^{-1}$ &  $5.41712\times10^{-3}$ & 1.62464 \\
\hline
$10^{-2}$ & $2.72990\times10^{1}$  & $4.43385\times10^{-2}$ & $3.85001\times10^{-1}$ & $1.15165\times10^{-1}$ & $5.67263\times10^{-1}$ & $3.75354\times10^{-1}$ & $2.41939\times10^{0}$ & $2.65677\times10^{-1}$ & $5.73964\times10^{-1}$ & $1.80675\times10^{-2}$ & 1.62527 \\
$10^{-3}$ & $2.65250\times10^{2}$  & $4.40157\times10^{-2}$ & $3.60001\times10^{-1}$ & $1.22266\times10^{-1}$ & $5.65797\times10^{-1}$ & $3.60993\times10^{-1}$ & $2.42868\times10^{0}$ & $2.66710\times10^{-1}$ & $5.73977\times10^{-1}$ & $5.72791\times10^{-3}$ & 1.62534 \\
$10^{-4}$ & $2.64430\times10^{3}$  & $4.40119\times10^{-2}$ & $3.57001\times10^{-1}$ & $1.23282\times10^{-1}$ & $5.65794\times10^{-1}$ & $3.59392\times10^{-1}$ & 
$2.43112\times10^{0}$ & $2.66740\times10^{-1}$ & $5.73722\times10^{-1}$ &  $1.59381\times10^{-2}$ & 1.62535 \\
$10^{-5}$ & $2.64335\times10^{4}$  & $4.40110\times10^{-2}$ & $3.57001\times10^{-1}$ & $1.23280\times10^{-1}$ & $5.65781\times10^{-1}$ & $3.59261\times10^{-1}$ & 
$2.42885\times10^{0}$ & $2.66731\times10^{-1}$ & $5.73980\times10^{-1}$ & $1.28157\times10^{-2}$ & 1.62534 \\
[0.5ex]
  \hline\hline
 \end{tabular}
 }
 
 \label{table:FDp0pt5}
\end{table}

\begin{table}
  \caption{Marginally stable configurations at different boundary temperatures $b = k_B T(R) / m c^2$ for a Fermi–Dirac distribution with $\alpha(R) = 1$, corresponding to an overly degenerate core.
  }
 \centering
 %\begin{ruledtabular}
 \resizebox{\columnwidth}{!}{
 \begin{tabular}{l l l l l l l l l l l l}
  \hline\hline
  % inserts table 
  %heading
  $b$  &  $w(0)$  & $\hat{M}$ & $\hat{R}$ & $C=\hat{M}/\hat{R}$ & $Z(0)$ & $\epsilon_c(0)/mc^2$ & $\hat{\rho}(0)$ & $\hat{p}(0)$ & $v(0)/c$ & $v(R)/c$ & $\langle\gamma\rangle=\gamma_{\text{cr}}$ \\ [0.5ex] 
  \hline\hline
$5.0$ & $6.46050\times10^{-2}$  & $4.02755\times10^{-1}$ & $5.03700\times10^{0}$ & $7.99592\times10^{-2}$ & $6.11611\times10^{-1}$ & $4.77152\times10^{-1}$ & $4.15454\times10^{-2}$ & $4.43410\times10^{-3}$ & $5.65850\times10^{-1}$ & $3.53112\times10^{-3}$ & 1.62361 \\
$3.0$ & $1.07604\times10^{-1}$  & $3.12882\times10^{-1}$ & $3.90600\times10^{0}$ & $8.01029\times10^{-2}$ & $6.11366\times10^{-1}$ & $4.76684\times10^{-1}$ & $6.87023\times10^{-2}$ & $7.33400\times10^{-3}$ & $5.65908\times10^{-1}$ & $4.49437\times10^{-3}$ & 1.62362 \\
$2.0$ & $1.61260\times10^{-1}$  & $2.56419\times10^{-1}$ & $3.19500\times10^{0}$ & $8.02562\times10^{-2}$ & $6.10998\times10^{-1}$ & $4.76057\times10^{-1}$ & $1.02011\times10^{-1}$ & $1.08918\times10^{-2}$ & $5.65960\times10^{-1}$ & $1.65126\times10^{-3}$ & 1.62363 \\
$1.0$ & $3.21650\times10^{-1}$  & $1.83381\times10^{-1}$ & $2.27000\times10^{0}$ & $8.07845\times10^{-2}$ & $6.09944\times10^{-1}$ & $4.74163\times10^{-1}$ & $1.97820\times10^{-1}$ & $2.11365\times10^{-2}$ & $5.66163\times10^{-1}$ & $1.82056\times10^{-3}$ & 1.62366 \\
\hline
$0.5$ & $6.39610\times10^{-1}$  & $1.32760\times10^{-1}$ & $1.62100\times10^{0}$ & $8.19001\times10^{-2}$ & $6.07691\times10^{-1}$ & $4.70157\times10^{-1}$ & $3.70892\times10^{-1}$ & $3.96896\times10^{-2}$ & $5.66598\times10^{-1}$ & $4.19084\times10^{-3}$ & 1.62373 \\
$0.3$ & $1.05751\times10^{0}$  & $1.06243\times10^{-1}$ & $1.27300\times10^{0}$ & $8.34589\times10^{-2}$ & $6.04620\times10^{-1}$ & $4.64647\times10^{-1}$ & $5.65332\times10^{-1}$ & $6.06340\times10^{-2}$ & $5.67240\times10^{-1}$ & $6.98121\times10^{-3}$ & 1.62382 \\
$0.2$ & $1.57027\times10^{0}$  & $9.04316\times10^{-2}$ & $1.05900\times10^{0}$ & $8.53933\times10^{-2}$ & $6.00850\times10^{-1}$ & $4.57813\times10^{-1}$ & $7.57414\times10^{-1}$ & $8.14694\times10^{-2}$ & $5.68056\times10^{-1}$ & $7.20648\times10^{-3}$ & 1.62394 \\
\hline
$0.15$ & $2.07315\times10^{0}$  & $8.16331\times10^{-2}$ & $9.35001\times10^{-1}$ & $8.73080\times10^{-2}$ & $5.97334\times10^{-1}$ & $4.51280\times10^{-1}$ & $9.03894\times10^{-1}$ & $9.74910\times10^{-2}$ & $5.68832\times10^{-1}$ & $8.85254\times10^{-3}$ & 1.62406 \\
$0.14$ & $2.21515\times10^{0}$  & $7.97954\times10^{-2}$ & $9.09001\times10^{-1}$ & $8.77836\times10^{-2}$ & $5.96378\times10^{-1}$ & $4.49515\times10^{-1}$ & $9.38887\times10^{-1}$ & $1.01338\times10^{-1}$ & $5.69037\times10^{-1}$ & $5.19705\times10^{-3}$ & 1.62409 \\
$0.13$ & $2.37820\times10^{0}$  & $7.79315\times10^{-2}$ & $8.82001\times10^{-1}$ & $8.83576\times10^{-2}$ & $5.95328\times10^{-1}$ & $4.47524\times10^{-1}$ & $9.76111\times10^{-1}$ & $1.05443\times10^{-1}$ & $5.69272\times10^{-1}$ & $1.66351\times10^{-3}$ & 1.62413 \\
$0.12$ & $2.56730\times10^{0}$  & $7.60441\times10^{-2}$ & $8.54001\times10^{-1}$ & $8.90445\times10^{-2}$ & $5.94136\times10^{-1}$ & $4.45242\times10^{-1}$ & $1.01552\times10^{0}$ & $1.09799\times10^{-1}$ & $5.69530\times10^{-1}$ & $2.74336\times10^{-3}$ & 1.62417 \\
$0.11$ & $2.78940\times10^{0}$  & $7.41345\times10^{-2}$ & $8.25001\times10^{-1}$ & $8.98599\times10^{-2}$ & $5.92805\times10^{-1}$ & $4.42637\times10^{-1}$ & $1.05722\times10^{0}$ & $1.14426\times10^{-1}$ & $5.69823\times10^{-1}$ &  $6.11333\times10^{-3}$ & 1.62422 \\
\hline
$0.1$ & $3.05400\times10^{0}$  & $7.22071\times10^{-2}$ & $7.95001\times10^{-1}$ & $9.08264\times10^{-2}$ & $5.91289\times10^{-1}$ & $4.39630\times10^{-1}$ & $1.10108\times10^{0}$ & $1.19309\times10^{-1}$ & $5.70146\times10^{-1}$ & $9.65430\times10^{-3}$ & 1.62427 \\
\hline
$0.09$ & $3.37490\times10^{0}$  & $7.02665\times10^{-2}$ & $7.65001\times10^{-1}$ & $9.18516\times10^{-2}$ & $5.89579\times10^{-1}$ & $4.36224\times10^{-1}$ & $1.14701\times10^{0}$ & $1.24444\times10^{-1}$ & $5.70511\times10^{-1}$ &  $6.72757\times10^{-3}$ & 1.62433 \\
$0.08$ & $3.77260\times10^{0}$  & $6.83210\times10^{-2}$ & $7.34001\times10^{-1}$ & $9.30802\times10^{-2}$ & $5.87648\times10^{-1}$ & $4.32269\times10^{-1}$ & $1.19468\times10^{0}$ & $1.29804\times10^{-1}$ & $5.70925\times10^{-1}$ & $1.73542\times10^{-3}$ & 1.62440 \\
$0.05$ & $5.87160\times10^{0}$  & $6.26124\times10^{-2}$ & $6.34001\times10^{-1}$ & $9.87576\times10^{-2}$ & $5.80093\times10^{-1}$ & $4.15554\times10^{-1}$ & $1.33967\times10^{0}$ & $1.46309\times10^{-1}$ & $5.72398\times10^{-1}$ & $8.43759\times10^{-3}$ & 1.62470 \\
\hline
$10^{-2}$ & $2.72360\times10^{1}$  & $5.69005\times10^{-2}$ & $4.92001\times10^{-1}$ & $1.15651\times10^{-1}$ & $5.67146\times10^{-1}$ & $3.74223\times10^{-1}$ & $1.46781\times10^{0}$ & $1.61181\times10^{-1}$ & $5.73963\times10^{-1}$ &  $1.37775\times10^{-2}$ & 1.62528 \\
$10^{-3}$ & $2.65180\times10^{2}$  & $5.65170\times10^{-2}$ & $4.62001\times10^{-1}$ & $1.22331\times10^{-1}$ & $5.65787\times10^{-1}$ & $3.60864\times10^{-1}$ & $1.47302\times10^{0}$ & $1.61760\times10^{-1}$ & $5.73973\times10^{-1}$ & $5.57449\times10^{-3}$ & 1.62534 \\
$10^{-4}$ & $2.64430\times10^{3}$  & $5.65117\times10^{-2}$ & $4.59001\times10^{-1}$ & $1.23119\times10^{-1}$ & $5.65805\times10^{-1}$ & $3.59472\times10^{-1}$ & 
$1.47455\times10^{0}$ & $1.61801\times10^{-1}$ & $5.73749\times10^{-1}$ & $6.52385\times10^{-3}$ & 1.62534 \\
$10^{-5}$ & $2.64335\times10^{4}$  & $5.65113\times10^{-2}$ & $4.58001\times10^{-1}$ & $1.23387\times10^{-1}$ & $5.65782\times10^{-1}$ & $3.59210\times10^{-1}$ & 
$1.47317\times10^{0}$ & $1.61780\times10^{-1}$ & $5.73980\times10^{-1}$ & $1.82917\times10^{-2}$ & 1.62534 \\
[0.5ex]
  \hline\hline
 \end{tabular}
 }
 
 \label{table:FDp1}
\end{table}

\end{document}